\documentclass[sigconf]{acmart}

\usepackage{booktabs} 
\usepackage{amsmath}
\usepackage{amsthm}
\usepackage{amssymb}
\usepackage{bm}
\usepackage{graphicx}
\usepackage{subfigure}
\usepackage{url}
\usepackage{enumitem}
\usepackage{bookmark}

\DeclareMathOperator{\diag}{diag}

\newcommand{\x}{\mathbf{x}}
\newcommand{\xx}{\mathbf{\tilde{x}}}
\newcommand{\xxx}{\mathbf{\hat{x}}}
\newcommand{\y}{\mathbf{y}}

\newcommand{\yyy}{\mathbf{\hat{y}}}
\newcommand{\uu}{\mathbf{u}}
\newcommand{\uuu}{\mathbf{\hat{u}}}
\newcommand{\vv}{\mathbf{v}}
\newcommand{\vvv}{\mathbf{\hat{v}}}

\newcommand{\Ex}{\mathbb{E}}
\newcommand{\pr}{\mathbb{P}}

\newcommand{\rt}{\right}
\newcommand{\lt}{\left}

\newcommand{\I}{\mathbf{I}}
\newcommand{\A}{\mathbf{A}}
\newcommand{\W}{\mathbf{W}}
\newcommand{\D}{\mathbf{D}}
\newcommand{\GG}{\mathcal{G}}
\newcommand{\Ii}{\mathcal{I}}
\newcommand{\Ss}{\mathcal{S}}
\newcommand{\N}{\mathcal{N}}
\newcommand{\E}{\mathcal{E}}

\newcommand{\Rn}{\mathbb{R}}

\theoremstyle{remark}
\newtheorem{remark}{Remark}

\allowdisplaybreaks

\setcopyright{none}

\begin{document}
\title{Transient Dynamics of Epidemic Spreading and Its Mitigation on Large Networks}

\author{Chul-Ho Lee}
\affiliation{%
  \institution{Florida Institute of Technology}
}
\email{clee@fit.edu}

\author{Srinivas Tenneti}
\affiliation{%
  \institution{Cisco Systems and North Carolina State University}
}
\email{stennet2@ncsu.edu}

\author{Do Young Eun}
\affiliation{%
  \institution{North Carolina State University}
}
\email{dyeun@ncsu.edu}

\begin{abstract}
In this paper, we aim to understand the transient dynamics of a susceptible-infected (SI) epidemic spreading process on a large network. The SI model has been largely overlooked in the literature, while it is naturally a better fit for modeling the malware propagation in early times when patches/vaccines are not available, or over a wider range of timescales when massive patching is practically infeasible. Nonetheless, its analysis is simply non-trivial, as its important dynamics are all transient and the usual stability/steady-state analysis no longer applies. To this end, we develop a theoretical framework that allows us to obtain an accurate closed-form approximate solution to the original SI dynamics on any arbitrary network, which captures the temporal dynamics over all time and is tighter than the existing approximation, and also to provide a new interpretation via reliability theory. As its applications, we further develop vaccination policies with or without knowledge of already-infected nodes, to mitigate the future epidemic spreading to the extent possible, and demonstrate their effectiveness through numerical simulations.
\end{abstract}

%
%



\maketitle

\setlength{\abovedisplayskip}{5pt}
\setlength{\belowdisplayskip}{5pt}

\section{Introduction}

\subsection{Background}
There has been an explosive growth in the number of the Internet-connected devices, ranging from traditional PCs to increasingly prevalent mobile devices and to recently popular Internet of things (IoT) devices. The number of connected devices on the Internet will exceed 50 billion by 2020, according to Cisco~\cite{cisco-forecast}. Apart from sheer volume, the end-device users have built a stack of rich and complex networks, such as online social networks (OSNs) and the networks of their contacts for calls and text messaging services like SMS/MMS and in the instant messaging applications, e.g., Skype and Google Hangouts, derived from their social, personal and work groups through which information can be shared and spread at an unprecedented rate. The prolific connections to end devices and users through potentially different networks clearly increase the risk of being exposed to malware and worm attacks and can also be exploited as devastating vehicles for their propagation.

For example, it has been recently demonstrated in~\cite{blackhat,Ronen-SP17,Morgner-WiSec17} that a Zigbee-enabled IoT device can be reprogrammed and compromised by an outside attacker in its vicinity, e.g., a Zigbee radio transceiver mounted on a drone, by exploiting the Zigbee's over-the-air firmware update mechanism in combination with the global firmware signing key obtained. The compromised device can further self-propagate a worm to nearby devices via the Zigbee wireless connectivity and thus potentially comprise the entire network. In a similar vein, by exploiting a buffer overflow vulnerability, it was shown that malware can be crafted to become a self-replicating worm over the air and the malware propagates over the network in a hop by hop manner~\cite{Giannetsos10,Giannetsos13}. In fact, the emergence of the over-the-air reprogramming protocols such as Deluge~\cite{Hui-SenSys04} and Trickle~\cite{Levis-NSDI04} to reconfigure a sensor network or disseminate code updates across the sensor network already gave rise to the potential misuse of its broadcasting nature to transfer malware across the whole sensor network quickly~\cite{Das-TMC09,Das-TOSN09}.

In addition, millions of OSN users have fallen prey to several malware/worm attacks, since the trust relationships established between OSN users can easily be exploited to spread worms~\cite{Samy,xss-worm,Koobface,Clickjacking,Yan-ASIACCS11,Faghani13}. For instance, the Samy worm, found in October 2005, exploited a cross-site scripting vulnerability against MySpace and infected over one million users across the MySpace within just 20 hours, by making any users viewing infected users' profiles infected and infectious to others~\cite{Samy}. The Koobface and Clickjacking worms, impacting Facebook in 2009 and 2010, respectively, leveraged the trust relationships among OSN users to make them click on links to malicious sites without much doubt and get infected to automatically share the links with their friends~\cite{Koobface,Clickjacking}. Exploiting such trust relationships between users can be traced to any kind of the existing communications between them, e.g., emails~\cite{Zou-TDSC07,Gao11}, text messages~\cite{Zhu-Infocom09,Gao13}, and instant messages~\cite{Rodpicom,Skype13} to propagate worm viruses through the networks of contacts and maximize their reachability to end hosts.

\vspace{-1mm}
\subsection{Motivation}

Since exploiting the network connectivity lies at the heart of the malware distribution, it becomes crucial to understand how the underlying network structure affects the malware propagation over the network. This helps us understand its spreading dynamics and eventually devise strategies for combating the malware propagation. Modeling the malware propagation and understanding its non-trivial properties have been active and important research topics in many disciplines. In the literature, most of the research efforts have been centered around the so-called epidemic threshold for the extinction of an epidemic under the Susceptible-Infected-Susceptible/Removed (SIS/SIR) models (and similar variants)~\cite{Wang03,Chakrabarti08,Ganesh05,Draief-AAP08,Mieghem-ToN09,Prakash-ICDM11,Ruhi-ArXiv16,Mei17}, where a susceptible node becomes infected with some constant rate $\beta$ and an infected node is assumed to be independently cured/removed (on its own) with some other constant rate $\delta$.\footnote{In the SIS model, any infected node is cured with $\delta$ and again becomes healthy and susceptible to infection, but in the SIR model, any infected node, once recovered, becomes completely cured and no longer susceptible to infection. The former captures the cases that variants of a worm are quickly developed one after another or multiple worms are concurrently present to continuously threaten every node, while the latter is suitable for the propagation of a worm over the network.} The epidemic threshold here indicates whether the ratio of the infection rate $\beta$ to the curing rate $\delta$, or the `effective' infection rate $\beta/\delta$, is greater or less than the reciprocal of a global network parameter $\lambda(\A)$ -- the largest eigenvalue of the adjacency matrix $\A$, which captures the structure of the underlying network.

A phase transition appears at the epidemic threshold below which the epidemic dies out eventually over time (so the network becomes virus-free in the steady state), but above which a non-zero fraction of nodes remain infected. Due to this dichotomic behavior, the `below-the-threshold' has been considered as the one and only condition to be achieved in order to eradicate the epidemic, and thus it has been the fundamental basis for the development of immunization strategies or the control of epidemics on a network~\cite{Holme-PRE02,Schneider-PRE11,Tong-TKDM16,Mieghem-PRE11,Vullikanti-SDM15,Tong-TKDD16,Preciado-CDC13,Preciado-TCNS14}. Specifically, the defense mechanisms have been developed, in essence, to ensure that the curing/recovery force across the network outweighs that of the infection, whose spreading is further fueled by the connectivity of the underlying network, so that the epidemic eventually dies out. It is achieved by either increasing the curing rate $\delta$ (possibly heterogeneous over different nodes) or altering/pruning the network connectivity for smaller $\lambda(\A)$ to disturb the epidemic spreading over the network.

All these rich results, however, become no longer meaningful when the patches or vaccines are not available from the beginning. It would rather make more sense to understand the properties of the Susceptible-Infected (SI) model that lacks the curing process, which captures the dynamics of an epidemic in early times when the recovery is not ready yet. Nonetheless, it has received much less attention from the research community on epidemic modeling and analysis~\cite{Canright06,Newman10,Mei17}, since every one eventually gets infected and remains infected, which may be trivial in view of the steady-state behavior and its analysis. In fact, the current literature has been mostly limited to the steady states of epidemics propagating over networks for the sake of tractability by resorting to the well-established steady-state/stability analysis tools. However, for the SI model, one would have to deal with the whole transient dynamics over \emph{all} time $t$, not just the steady-state.

\vspace{-0mm}
\subsection{Contributions}

In this paper, we develop a theoretical framework to characterize the `transient' dynamics of the SI epidemic dynamics on \emph{any arbitrary} network. We first revisit the standard SI model on a network and discuss the limitations of the current literature. We then propose a simple yet effective technique to obtain an accurate closed-form approximate solution to the SI epidemic dynamics over all time $t$, which effectively overcomes the limitations and becomes tighter than the existing linearized approximation~\cite{Canright06,Newman10,Mei17}. The rational behind the technique is to judiciously transform the dynamical system governing the SI dynamics into an equivalent system, which makes mathematical analysis amenable. Our transformation also helps discover an equivalent interpretation of the SI dynamics from the viewpoint of reliability theory, which is concerned with how much longer each node would survive given its current status. Leveraging our theoretical findings, we finally develop vaccination policies on how to distribute a limited amount of patches or vaccines over the network in order to preventively mitigate any future potential attacks and also to reactively minimize the aftermath of an epidemic spread that already took place. We evaluate our vaccination policies against the eigenvector centrality-based policy, which is backed up by the conventional linearized SI model~\cite{Canright06,Carreras2007,Newman10,Lu-PR16}, and the degree-based policy over real network topology datasets, to demonstrate the efficacy of our policies.

\section{Preliminaries}\label{se:prelim}

\subsection{Standard Epidemic Models on Networks}

Consider a general connected, aperiodic, undirected graph $\GG \!=\! (\N, \E)$ to model a network with a set of nodes $\N \!=\! \{1, 2, \ldots , n\}$ and a set of edges $\E$, indicating neighboring relationships between nodes. The graph $\GG$ is defined by an $n \times n$ adjacency matrix $\A \!=\! [a_{ij}]$ with elements $a_{ij} \!=\! 1$ if there is an edge between nodes $i$ and $j$, i.e., $(i,j) \!\in\! \E$, and $a_{ij} \!=\! 0$ if otherwise.\footnote{For ease of exposition, we focus on the unweighted graph in this paper, although our theoretical results can be generalized to the weighted graph.} Since $\GG$ is connected and undirected, $\A$ is irreducible, i.e., for any nodes $i$ and $j$ there exists some integer $k$ such that $(\A^k)_{ij} \!>\! 0$, and symmetric, i.e., $a_{ij} \!=\! a_{ji}$ for all $i,j$~\cite{Meyer00}. Similarly for the aperiodicity of $\A$. In the standard epidemic models over a graph $\GG$, the network connectivity mainly constraints the infection process, i.e., if node $i$ is infected and node $j$ is susceptible for $(i,j) \!\in\! \E$, then $i$ can spread an undesired virus or simply infect $j$ with a common infection rate $\beta \!>\! 0$. Sharing this common network-constrained infection process, various epidemic models can be related to each other.

In the standard Susceptible-Infected-Susceptible (SIS) epidemic model over an undirected graph with $\A$, each susceptible node becomes infected at the infection rate $\beta$ \emph{per link} times the number of infected neighboring nodes, while any infected node is independently cured on its own with a curing rate $\delta \!>\! 0$ and becomes healthy (again susceptible to infection). The SIS model can also be extended to the Susceptible-Infected-Removed (SIR) model with only difference that a node, once recovered with rate $\delta$, becomes completely cured and no longer susceptible to infection.
These models turn into the Susceptible-Infected (SI) model, if there is no curing process (i.e., $\delta \!=\! 0$), implying that a node, once infected, stays infected forever and does not fall back to the susceptible/removed state.

Let $X_i(t) \!\in\! \{0,1\}$ denote the state of node $i$ at time $t$, where $X_i(t) \!=\! 1$ indicates that node $i$ is infected and $X_i(t) \!=\! 0$ indicates that node $i$ is healthy and susceptible to infection at time $t$. Define $x_i(t) \!\triangleq\! \pr\{X_i(t) \!=\! 1\} \!=\! \Ex\{X_i(t)\} \!\in\! [0,1]$ to be the probability that node $i$ is infected at time $t$. In other words, node $i$ is healthy at time $t$ with probability $1 \!-\! x_i(t)$. Let $\x(t) \!\triangleq\! [x_1(t), x_2(t), \ldots, x_n(t)]^T$ be an $n$-dimensional column vector with elements $x_i(t)$ at time $t$. Then, the SIS model is characterized by the following system of $n$ coupled, nonlinear differential equations~\cite{Newman10,Mieghem-ToN09,Mieghem-PRE12,Nowzari-CS16}: For any node $i \in \N$,
\begin{equation}
  \frac{d x_i(t)}{dt} = \beta (1-x_i(t)) \sum_{j\in\N} a_{ij} x_j(t)  - \delta x_i(t), \quad t \geq t_0 \label{basic-SIS}
\end{equation}
with a pre-specified initial condition $\x(t_0)$. Note that there are other epidemic models. However, the traditional ones, e.g., compartmental model and metapopulation model, neglect the underlying network structure and assume the \emph{homogeneous mixing} population~\cite{Sarkar-TAC15,Sarkar-ToN16}, in that every individual has equal chance to contact others in the population or there are multiple homogeneous subgroups, which are clearly far from reality. There have also been other epidemic `network' models based on the degree-based approximation~\cite{Pastor-Satorras-PRL01,Pastor-Satorras-PRE01,Newman10,Kuri-ToN16}, where the SIS epidemic process is defined on the so-called configuration model, or a random graph with a given degree distribution. It has, however, been shown in~\cite{Zou-TDSC07,Mieghem-PRE12} that the degree-based approximation is inaccurate, because the underlying network is implicitly assumed to be tree-like and uncorrelated over degrees, thus implying that all nodes with the same degree are indistinguishable regardless of how they are connected to the network.

The SIS model in (\ref{basic-SIS}) exhibits the existence of the epidemic threshold $\tau_c \!=\! 1/\lambda(\A)$, where $\lambda(\A)$ is the largest eigenvalue of $\A$, i.e., the spectral radius of $\A$. If the `effective' infection rate $\beta/\delta$ satisfies $\beta/\delta \! < \! \tau_c$, the epidemic dies out over time, i.e., $x_i(\infty) \!=\! 0$ for all $i$. However, if $\beta/\delta \! > \! \tau_c$, a non-zero fraction of infected nodes persist, $\frac{1}{n} \sum_{i=1}^{n} x_i(\infty) \!>\! 0$. The epidemic threshold was originally shown in~\cite{Wang03,Chakrabarti08} under a discrete-time version of (\ref{basic-SIS}). This has been made precise rigorously via a stochastic analysis~\cite{Ganesh05} and using mean-field theory~\cite{Mieghem-ToN09}.\footnote{Strictly speaking, the exact model is a sequence of $(X_1(t), \ldots, X_n(t))$, which is an absorbing Markov chain with $2^n$ states having an absorbing state $\bm{0}$. Since this chain is `finite' and irreducible on transient states, the steady state is the absorbing state, i.e., the extinction of the epidemic. Nonetheless, the epidemic threshold $\tau_c$ still plays a critical role in determining whether the epidemic dies out fast or lasts long. For the latter case, a non-zero fraction of infected nodes persist over a long time span of practical interest, which is called the `metastable' state. This corresponds to the case under the exact Markovian model, in which the steady state $\bm{0}$ is only reached after an unrealistically long time, i.e., it is in reality never reached~\cite{Ganesh05,Mieghem-ToN09,Mieghem-PRE12}. The SIS model in (\ref{basic-SIS}) is also known to be very accurate with large $n$~\cite{Mieghem-ToN09,Mieghem-PRE12}.} Similar observation has also been made for the SIR model~\cite{Draief-AAP08}. Since then, there have been a great deal of follow-up theoretical studies regarding the epidemic threshold with its extension to different epidemic models such as SIRS, SEIR and SIV  models~\cite{Prakash-ICDM11,Ruhi-ArXiv16,Mei17}.\footnote{SIRS, SEIR and SIV stand for susceptible-infected-recovered-susceptible, susceptible-exposed-infected-recovered, and susceptible-infected-vaccinated, respectively.}

The epidemic threshold $\tau_c$ has been the cornerstone of the control of epidemics on a network. The development of immunization strategies or the control of epidemics on a network has been mostly made to ensure the `below-the-threshold' condition $\beta/\delta \! < \! \tau_c$ for the (quick) extinction of an epidemic by either decreasing $\beta/\delta$ or $\lambda(\A)$. For example, Many prior efforts have been made to minimize $\lambda(\A)$ in order to ensure the extinction of an epidemic by removing a set of $k$ nodes~\cite{Holme-PRE02,Schneider-PRE11,Tong-TKDM16} or $k$ edges~\cite{Mieghem-PRE11,Vullikanti-SDM15,Tong-TKDD16} from the network. Since the node and edge removal problems are generally known to be NP-complete~\cite{Mieghem-PRE11}, heuristic algorithms have been proposed for the problems based on their own way of measuring the importance of nodes and/or edges~\cite{Holme-PRE02,Schneider-PRE11,Tong-TKDM16,Mieghem-PRE11,Vullikanti-SDM15,Tong-TKDD16}. Another class of approaches has been to allow heterogeneous $\delta_i$ and $\beta_i$ over nodes $i$ and find the optimal $\delta^*_i$ and/or $\beta^*_i$ to minimize a modified form of $\lambda(\A)$ under various constraints, yet still to ensure the extinction of an epidemic under the SIS/SIR models~\cite{Preciado-CDC13,Preciado-TCNS14,Nowzari-CS16}.

\vspace{-1mm}
\subsection{Notations}
We present notations that will be used throughout the rest of the paper. For any two column vectors $\uu \!=\! [u_1, u_2, \ldots, u_n]^T \!\in\! \Rn^n$ and $\vv \!=\! [v_1, v_2, \ldots, v_n]^T \!\in\! \Rn^n$, we write
\begin{equation*}
  \uu \preceq \vv \;\; \text{if} \;\; u_i \leq v_i \;\; \text{for all} \; i = 1,2 \ldots, n.
\end{equation*}
Similarly, we write $\uu \!\prec\! \vv$ if $u_i \!<\! v_i$ for all $i$. Note that this componentwise inequality between vectors is a partial order, which satisfies reflexivity, anti-symmetry, and transitivity. We denote by $\bm{1}$ the $n$-dimensional column vector whose elements are all ones. Similarly for $\bm{0}$ with all-zero elements. With a slight abuse of notation, for a function $f$: $\Rn \!\to\! \Rn$ and for a column vector $\uu \!\in\! \Rn^n$, we write $f(\uu)$ as a $n$-dimensional column vector with elements $f(u_i)$. Similarly, let $\diag(\uu)$ be an $n \times n$ diagonal matrix with diagonal entries $u_i$. Let $\I \!=\! \diag(\bm{1})$ be an $n \times n$ identity matrix. We adopt the standard convention $0 \log(0) = \lim_{x \to 0} x \log(x) = 0$ in this paper.

\vspace{-0mm}
\section{Motivations for the SI Model}\label{se:motiv}

We present, from practical perspectives, why the SI model is most relevant in early times when patches/vaccines are not available, or possibly over a wider range of timescales under constrained environments where applying massive patches to end hosts is practically infeasible.

\vspace{-1mm}
\subsection{Practical Implications of the Epidemic Threshold}

We examine how the epidemic threshold $\tau_c$ plays in real network setting. For any graph $\GG$, we observe that
\begin{equation*}
\setlength{\abovedisplayskip}{4pt}
\setlength{\belowdisplayskip}{4pt}
  \max\{\bar{d}, \sqrt{d_{\max}}\} \leq \lambda(\A) \leq d_{\max},
\end{equation*}
where $\bar{d}$ is the average degree and $d_{\max}$ is the maximum degree~\cite{Lovasz07,Newman10}. This implies that the curing rate $\delta$ must be at least $\max\{\bar{d}, \sqrt{d_{\max}}\}$ times greater than the infection rate $\beta$ in order to ensure the extinction of an epidemic. Considering the million- to billion-scale networks currently prevalent in reality that often possess highly skewed degree distributions, the square root of the maximum degree $\sqrt{d_{\max}}$ is generally very large, although the average degree $\bar{d}$ can be small. Thus, we expect that the maximum eigenvalue $\lambda(\A)$ tends to be very large, so is the required $\delta/\beta \!\geq\! \lambda(\A)$. Since applying the massive patches to nodes on a large-scale network is often infeasible in practice as shall be explained shortly, achieving the below-the-threshold condition by increasing $\delta$ (possibly $\delta_i$) and lowering $\lambda(\A)$ by removing a few nodes and/or edges can hardly be practical. In addition, we note that the immediate availability of the curing process (possibly with very high curing rate) makes the SIS/SIR models over-simplified and far from the reality, since any infected devices remain infected at least over the early times or possibly over a longer time span, depending on when the patches or workarounds for the underlying epidemic become available.

\begin{figure}[t!]
    \subfigcapskip = -1mm
    \centering
    \vspace{-0mm}
    \hspace{-10mm}\subfigure[]{\includegraphics[width=1.5in,height=1.3in]{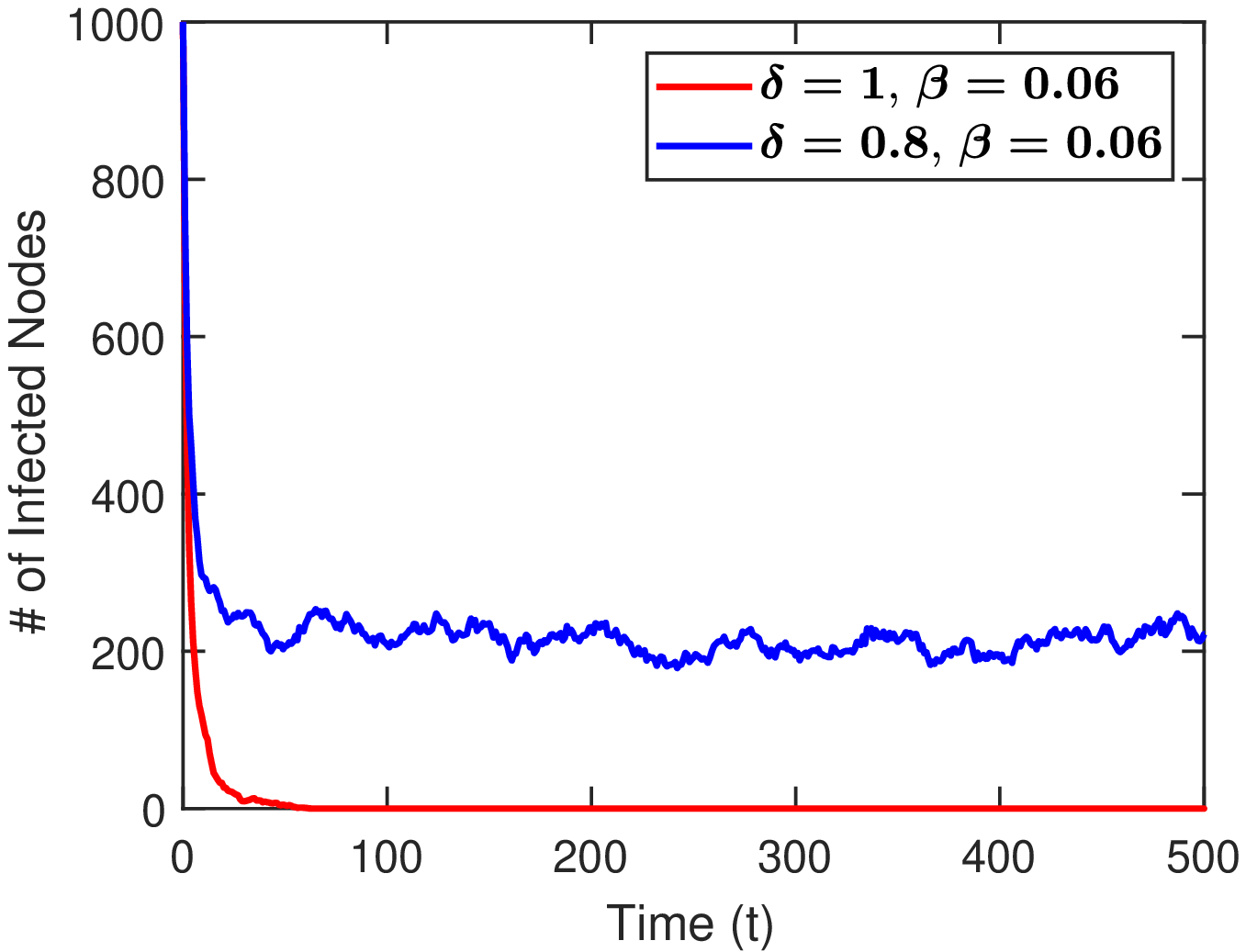}}
    \hspace{-0mm}\subfigure[]{\includegraphics[width=1.9in,height=1.25in]{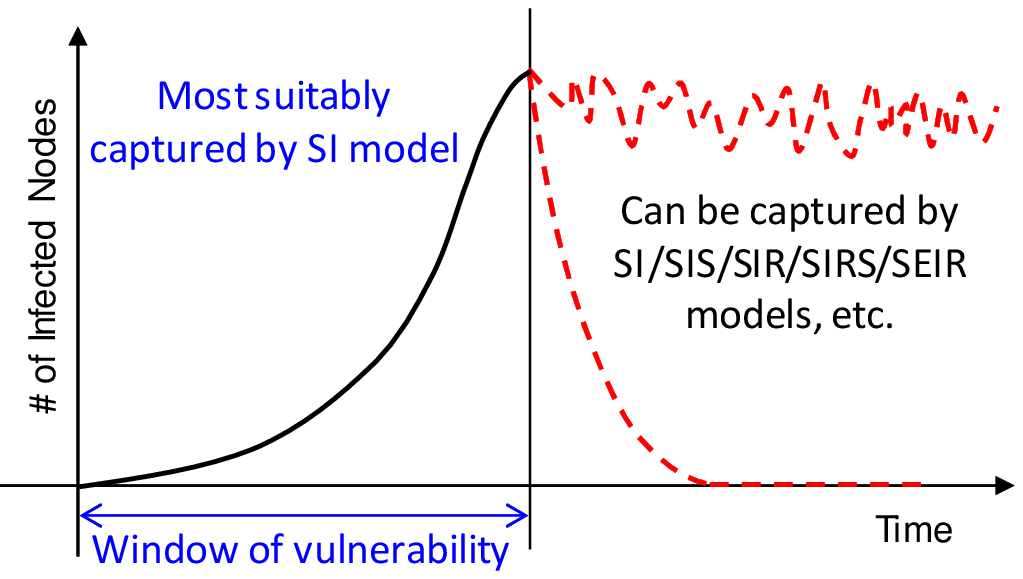}}
    \hspace{-12mm}
    \vspace{-4mm}
    \caption{\small (a) The dynamics of an SIS epidemic over an Erd\H{o}s-R\'{e}nyi graph; (b) Illustrating the epidemic evolution before and after the window of vulnerability.}\label{fig:sis}
    \vspace{-6mm}
\end{figure}

As an example, consider an Erd\H{o}s-R\'{e}nyi graph with $n \!=\! 2000$ and $\lambda(\A) \!=\! 16.159$, which is the same simulation setting as in~\cite{Ruhi-ArXiv16}. We present two simulation results on the evolution of an SIS epidemic for below and above the threshold in Figure~\ref{fig:sis}(a). For the SIS epidemic, we fix $\beta \!=\! 0.06$, and use $\delta \!=\! 1$ for below the threshold and $\delta \!=\! 0.8$ for above the threshold. As the initial condition, 1000 initially infected nodes are randomly selected. While Figure~\ref{fig:sis}(a) confirms that the epidemic dies out quickly for the former and a non-zero fraction of infected nodes remain in any observable time for the latter, we note that the extinction of the epidemic requires the curing rate $\delta$ to be at least $\lambda(\A) \!\approx\! 16$ times greater than the infection rate $\beta$ in this simulation, i.e., the infection itself is weaker than the autonomous self-recovery by $\lambda(\A) \!\approx\! 16$ times, which seems to be either too idealistic or practically infeasible in reality.

\vspace{-1mm}
\subsection{The Gap between Theory and Practice}\label{se:practical}

There is a huge gap between theory and practice. First, the patches and workarounds are not immediately available. There is often a non-negligible amount of time required for a patch or vaccine to be available when a vulnerability or its exploit becomes disclosed. In particular, when zero-day vulnerabilities are exploited, the attacks are rarely discovered right away. More often than not, there is an undisclosed vulnerability that may have been already exploited, but its patch is not available or the exploit even remains unknown to developers and engineers. The window of vulnerability between the time a threat is discovered and the time a security patch is released can take not just days but weeks or months and sometimes years~\cite{zeroday1,zeroday2}. Therefore, the early times for the malware/worm attacks are the non-negligible critical timescales for which the SI model is most suitable than any other models armed with the curing process, as illustrated in Figure~\ref{fig:sis}(b)\footnote{Figure~\ref{fig:sis}(b) also represents two possible dynamics after the window of vulnerability. One is the extinction of the underlying epidemic with massive patching, and the other is the presence of a non-negligible number of infected nodes, which happens when massive patching is practically not possible.}, but little is known for the transient dynamics of the SI model.

Second, patching end hosts often comes at a significant cost. Take applying patches to IoT devices as an example. The IoT devices generally adopt low-power wireless technologies supporting very low data rates as compared to WiFi and cellular networks~\cite{LPWAN}. They are also often deployed with duty-cycle constraints (the maximum duty cycle is typically 1\% or 0.1\%)~\cite{Zigbee,ETSI,Adelantado17}, which means that the devices are not connected to the network all the time. The low bandwidth is thus common in most IoT networks. In addition, sending a technician to patch manually is not possible due to the huge number of devices that need to be patched and because they may be located in harsh environments~\cite{OilandGasAttack}, making it difficult to directly access the devices and thus rendering the over-the-air updates the only viable solution. Considering the low bandwidth of the IoT networks, distributing patches, once available, to IoT devices over the air becomes a daunting, high-cost task both in time and bandwidth required. Thus, achieving the `below-the-threshold' for the extinction of an epidemic is deemed practically impossible over a wide range of timescales, since the difficulties of distributing patches tend to make the long-term recovery rate $\delta$ small and thus the effective infection rate $\beta/\delta$ above the threshold. In this above-the-threshold regime, the node infection is more dominant than its recovery and the corresponding epidemic dynamics can be more or less captured by the largely-overlooked SI model, which is in contrast to the existing literature that typically strives to achieve the complete extinction of an epidemic. Therefore, we would have to seek to alleviate the epidemic spreading to the maximum extent possible under the patching-cost constraints via the marginalized SI model, or the SIS/SIR models in the above-the-threshold regime.

\vspace{-0mm}
\section{Revisiting the SI Model on a Graph}\label{se:model}

We have observed that the SI model is most relevant in early times when patches are not available, or over a wider range of timescales when massive patching is practically infeasible. We first note that in contrast to the SIS/SIR models, the only \emph{fixed point} (\emph{steady state}) of this SI dynamics is $\x(\infty) \!=\! \bm{1}$ as long as $\x(t_0) \neq \bm{0}$, and thus it becomes necessary (and non-trivial) to understand the transient dynamics of $\x(t)$ for all $t$, not just the fixed point $\x(\infty)$. Unfortunately, there is not much known about the SI model, as opposed to the abundant literature on the SIS/SIR models that is mostly based on the stability/steady-state analysis regarding $\x(\infty)$.

By setting $\delta \!=\! 0$ in (\ref{basic-SIS}), we write again the dynamical system for the standard SI model as follows. For any node $i \in \N$,
\begin{equation}
  \frac{d x_i(t)}{dt} = \beta (1-x_i(t)) \sum_{j\in\N} a_{ij} x_j(t), \quad t \geq t_0, \label{basic-SI}
\end{equation}
with an initial condition $\x(t_0)$. To avoid the triviality, we exclude the trivial cases $\x(t_0) \!=\! \bm{0}$ and $\x(t_0) \!=\! \bm{1}$ throughout the paper.\footnote{If $\x(t_0) \!=\! \bm{0}$, every node is healthy at time $t_0$ with probability 1 and thus they remain healthy forever. Similarly, if $\x(t_0) \!=\! \bm{1}$, every one is already infected at time $t_0$.} Thus, there is at least one node $i$ having strictly positive probability of being infected at time $t_0$, i.e., $x_i(t_0) \!>\! 0$ for some $i$. Note that $x_i(t)$ is non-decreasing in time $t \!\geq\! t_0$ as the RHS of (\ref{basic-SI}) is non-negative for all $t$, for all $i$, i.e., $\x(t_1) \preceq \x(t_2)$ for $t_1 \leq t_2$.

The nonlinear dynamical system of the SI model in (\ref{basic-SI}), albeit its simple form, is generally not solvable in a closed form. Linearization has thus been used in the literature explicitly or implicitly as an approximation for the SI model~\cite{Canright06,Newman10,Mei17}. It is essentially to use a linear dynamical system that upper-bounds the original nonlinear dynamical system in (\ref{basic-SI}) as follows. For any $i \in \N$ and for any time $t \geq t_0$,
\begin{equation}
  \frac{d x_i(t)}{dt} =  \beta (1 - x_i(t)) \sum_{j\in\N} a_{ij} x_j(t)   \leq \beta \sum_{j\in\N} a_{ij} x_j(t) \label{SI-inequality}
\end{equation}
from $1 - x_i(t) \leq 1$. Let $\xx(t) \triangleq [\tilde{x}_1(t), \tilde{x}_2(t), \ldots, \tilde{x}_n(t)]^T$ be the solution of the following linear dynamics, which is available in a closed form. For any $t \geq t_0$,
\begin{equation*}\label{basic-SI-linear-matrix}
  \frac{d \tilde{x}_i(t)}{dt} =  \beta \sum_{j\in\N} a_{ij} \tilde{x}_j(t), \; \text{or} \;\; \frac{d \xx(t)}{dt} = \beta \A \xx(t).
\end{equation*}
It thus follows from (\ref{SI-inequality}) that
\begin{equation}
  \x(t) \preceq \xx(t) = e^{\beta  (t-t_0) \A}\x(t_0), \quad t \geq t_0, \label{basic-SI-linear-solution}
\end{equation}
given that $\xx(t_0) = \x(t_0)$. To the best of our knowledge, it is the only mathematical technique available in the literature for tractable analysis of the SI model. Note that the linearization has also been used for other epidemic models that have nonlinear system dynamics~\cite{Wang03,Chakrabarti08,Newman10,Preciado-CDC13,Preciado-TCNS14}.

$\xx(t)$ can also be expressed in terms of the eigenvalues and eigenvectors of $\A$. First, the spectral decomposition of the symmetric matrix $\A$ yields $\A \!=\! \sum_{k=1}^n \lambda_k \vv_k \vv_k^T$, where $\lambda_1 \!\geq\! \lambda_2 \!\geq\! \cdots \!\geq\! \lambda_n$ are the real eigenvalues of $\A$, and $\vv_i$ are orthonormal eigenvectors of $\A$ with $\vv_i^T \vv_i \!=\! 1$ for any $i$, and $\vv_i^T \vv_j \!=\! 0$ for $i \!\neq\! j$~\cite{Meyer00}. Then, by representing $e^{\A t}$ as a power series, we have
\vspace{-0.5mm}
\begin{align*}
  e^{\A t} &= \I + t\A  + \frac{t^2}{2!}\A^2 + \frac{t^3}{3!}\A^3 + \cdots \\
  &= \sum_{k=1}^n \vv_k \vv_k^T + t \sum_{k=1}^n \lambda_k \vv_k \vv_k^T + \frac{t^2}{2!}\sum_{k=1}^n \lambda^2_k \vv_k \vv_k^T + \cdots, \\
  &= \sum_{k=1}^n \vv_k \lt(1 + t\lambda_k + \frac{(t\lambda_k)^2}{2!} + \cdots \rt) \vv_k^T  = \sum_{k=1}^n e^{\lambda_k t} \vv_k \vv_k^T,
\end{align*}
where we have used the identity $\I = \sum_{k=1}^n \vv_k \vv_k^T$ and the orthonormality of $\vv_i$. Furthermore, since the non-negative matrix $\A$ is irreducible and aperiodic, by Perron-Frobenius theorem~\cite{Meyer00}, $\lambda_1 \!=\! \lambda(\A)$ is positive with $\lambda_1 \!>\! |\lambda_k|$ ($k \!\neq\! 1$) and its corresponding eigenvector $\vv_1$ has all positive components, i.e., $\vv_1 \!\succ\! \bm{0}$. Therefore, from (\ref{basic-SI-linear-solution}), we have, for any $t \geq t_0$,
\vspace{-1mm}
\begin{align}
  \xx(t) &= e^{\beta  (t-t_0) \A}\x(t_0) = \sum_{k=1}^n \xi_k e^{\beta \lambda_k (t-t_0)} \vv_k \nonumber \\
  &= e^{\beta \lambda_1 (t-t_0)} \lt( \xi_1 \vv_1 + \sum_{k=2}^n\xi_k e^{-\beta (\lambda_1 - \lambda_k) (t-t_0)} \vv_k \rt), \label{spectral} \\
  &= \xi_1 e^{\beta \lambda_1 (t-t_0)}\vv_1 \lt(1 + O\lt(e^{-\min_{k\geq 2} \!|\lambda_1 - \lambda_k|(t-t_0)}\rt)\rt), \label{EVC-approx}
\end{align}
where $\xi_k \!\triangleq\! \vv_k^T\x(t_0) \!\in\! \mathbb{R}$ for each $k$. By noting that $\lambda_1 \!>\! 0$, $\vv_1 \!\succ\! \bm{0}$, and $\xi_1 \!=\! \vv_1^T\x(t_0) \!>\! 0$, (\ref{EVC-approx}) follows from that all the summands in (\ref{spectral}) decay exponentially fast in time $t$, since $\lambda_1 \!-\! \lambda_k \!>\! 0$ for $k \!\geq\! 2$. The leading eigenvector $\vv_1$ is widely known as the eigenvector centrality (EVC)~\cite{Canright06,Carreras2007,Newman10,Lu-PR16}, of which $i$-th element represents the notion of importance (centrality) of node $i$. Thus, it has been argued that the EVC captures the \emph{initial} growth rate of the SI epidemic dynamics, or the probability of node $i$ being infected, over early time $t$~\cite{Canright06,Carreras2007,Newman10,Mei17}.

However, we observe that there are the following critical limitations with the `linearization bound' in (\ref{basic-SI-linear-solution}).
\vspace{-1mm}
\begin{itemize}[itemsep=-0.5pt,leftmargin=1.8em]
\item[1.] The linearization bound $\xx(t)$ in (\ref{basic-SI-linear-solution}) and its approximation via $\vv_1$ in (\ref{EVC-approx}) quickly grow \emph{without bound}, having $\tilde{x}_i(\infty) \!=\! \infty$ for all $i$. By definition, it must be $\x(t) \preceq \bm{1}$ for all $t$.

\item[2.] $\xx(t)$ is an accurate approximate solution to the original SI dynamics in (\ref{basic-SI}) \emph{only} for \emph{small} time $t$ and \emph{only} when $\x(t_0) \approx \bm{0}$.

\item[3.] The approximation of $\xx(t)$ via the EVC $\vv_1$ is only effective for \emph{large} time $t$ in which regime $\xx(t)$ already becomes invalid, rendering $\tilde{x}_i(t) \!\gg\! 1$, although the EVC has been mistakenly known to capture the \emph{early} infection dynamics.
\end{itemize}
\vspace{-1mm}
The first limitation is apparent because the dominant term in (\ref{EVC-approx}) grows exponentially fast in $t$, and the second one can be seen from that $\x(t_0) \approx \bm{0}$ implies $1 - x_i(t) \approx 1$ for all $i$ and thus makes the upper bound tight in (\ref{SI-inequality}) only over early $t$ near $t_0$.\footnote{The most likely scenario ensuring $\x(t_0) \!\approx\! \bm{0}$ would be that the fraction of the initial infective nodes to the graph size $n$ is small but there is no knowledge on the exact initial infectives. For example, a small number $c$ of nodes, which are chosen \emph{uniformly at random}, are initially infected, i.e., $x_i(t_0) \!=\! c/n $ for all $i$, when $n$ is large~\cite{Newman10}. However, $\x(t_0) \!\approx\! \bm{0}$ is no longer valid as long as the knowledge on the exact initial infectives is available. For instance, if a specific node, say $i^*$, is infected at $t_0$ and \emph{its identity is known}, then $x_{i^*}(t_0) \!=\! 1$ and $x_{i}(t_0) \!=\! 0$ for all other nodes $i$, making $\x(t_0) \!\approx\! \bm{0}$ invalid. That is, $\x(t_0) \!\approx\! \bm{0}$ does not always hold even when there are few infected nodes at $t_0$.} The third limitation follows from that the approximation via the EVC $\vv_1$ in (\ref{EVC-approx}) is close to $\xx(t)$ only when time $t$ is sufficiently large enough.

Therefore, the conventional use of the EVC~\cite{Canright06,Carreras2007,Newman10,Lu-PR16} to characterize the SI epidemic dynamics and its applications to identify critical nodes for vaccine allocation become largely questionable. This also calls for a more precise approximate solution of (\ref{basic-SI}) that works for any arbitrary $\x(t_0)$ over a longer period of time $t$, which will lead to more effective policies for combating the epidemics and providing vaccine distribution.

\vspace{-0mm}
\section{Understanding the Transient Dynamics of SI Epidemic Spreading}\label{se:transform}

Having observed the limitations of the conventional linearized ODE approach to the SI model, we set out to develop a simple yet effective technique to obtain a closed-form approximate solution to the original SI dynamics in (\ref{basic-SI}). In contrast to the conventional one, our solution is valid and remains effective for any initial condition $\x(t_0)$ and for all time $t$. In particular, ours is much tighter than the conventional one, as shall be demonstrated below. Our technique is based upon a novel transformation converting the dynamical system in (\ref{basic-SI}) into an equivalent system with a suitable substitution. This transformation allows us to obtain the closed-form approximate solution and, in turn, to better characterize the transient dynamics of the original SI epidemic spreading process. In addition, we newly uncover an equivalent interpretation of the SI dynamics from a reliability theory standpoint and explain its connection to our transformation. This also reveals the stochastic properties of how long each node $i$ would survive as time $t$ increases, thanks to the interpretation and transformation.

\vspace{-0mm}
\subsection{Transformation Method and Tight Bound}

Our key observation is to set
\begin{equation}\label{transform-y}
  y_i(t) \triangleq g(x_i(t)) = -\log(1-x_i(t)) \in [0,\infty],
\end{equation}
where $g(x) \!\triangleq\! -\log(1\!-\!x)$ is an increasing convex function in $x\!\in\![0,1]$. Equivalently, we have
\begin{equation}\label{transform-x}
  x_i(t) \triangleq f(y_i(t)) = 1 - e^{-y_i(t)} \in [0,1],
\end{equation}
where $f(y) \!\triangleq\! 1 \!-\! e^{-y} \!=\! g^{-1}(y)$ is the inverse of $g$, and increasing concave in $y \!\geq\! 0$. Thus, (\ref{basic-SI}) can be written as
\begin{equation}
  \frac{dy_i(t)}{dt} = \beta \sum_{j\in\N} a_{ij}f(y_j(t)), \quad t \geq t_0, \label{transform-SI}
\end{equation}
with $\y(t_0) \!=\! g(\x(t_0))$. By applying the `inverse' substitution with $f$, we can recover the original vector $\x(t) \!=\! f(\y(t))$ at time $t$. Note that in addition to the aforementioned monotonicity property over time $t$, i.e., $\x(t_1) \preceq \x(t_2)$ for any $t_1 \leq t_2$, we can see that there is another monotonicity over ordered initial conditions. Letting $\underline{\x}(t)$ be the solution of (\ref{basic-SI}) with a different initial condition $\underline{\x}(t_0)$, if $\x(t_0) \preceq \underline{\x}(t_0)$, then $\x(t) \preceq \underline{\x}(t)$ for all $t \geq t_0$. This can be easily seen from that $\y(t) \!=\! g(\x(t))$ is a monotonic transformation of $\x(t)$ (and vice versa), and the RHS of (\ref{transform-SI}) is non-decreasing in $y_i(t)$ for all $t \!\geq\! t_0$. It is worth noting that this property is not entirely obvious in the view of (\ref{basic-SI}) but becomes apparent through our transformation.

Similar to the original dynamics in (\ref{basic-SI}), unfortunately, its transformed dynamics in (\ref{transform-SI}) is still nonlinear and is generally not solvable in a closed form. While we are inevitably led to derive an upper bound of the (unknown) solution of (\ref{basic-SI}), it effectively overcomes the limitations of the conventional one in (\ref{basic-SI-linear-solution}) and remains valid for any initial conditions $\x(t_0)$ and for all $t \geq t_0$. We below show this upper bound.

For a given $\x(t_0)$, let $\xxx(t) \!\triangleq\! [\hat{x}_1(t), \hat{x}_2(t), \ldots, \hat{x}_n(t)]^T$ be our approximate solution to (\ref{basic-SI}), which is given by $\xxx(t) \!=\! f(\yyy(t))$, where $\yyy(t)$ is the solution of the following nonhomogeneous linear system and $\yyy(t_0) \!=\! g(\x(t_0))$: For any $t \geq t_0$,
\begin{equation}
  \frac{d\yyy(t)}{dt} = \beta \A \diag(\bm{1} \!-\! \x(t_0)) \yyy(t) + \beta \A b(\x(t_0)), \label{upper2}
\end{equation}
where $b(x) \triangleq x + (1 - x)\log(1 - x)$ is increasing from $b(0) = 0$ to $b(1) = 1$. We have the following.

\vspace{-0.5mm}
\begin{theorem}\label{thm:upper}
For any $t \!\geq\!  t_0$, we have
\begin{equation}
  \x(t) \,\preceq\, \xxx(t) = f(\yyy(t)) \,\preceq\, \xx(t), \label{ordering}
\end{equation}
when they have the same initial conditions, i.e., $\x(t_0) = \xxx(t_0) = \xx(t_0)$, where $\yyy(t)$ is given by
\begin{align}
  \yyy(t)  &= e^{\beta(t-t_0) \A\diag(\bm{1} - \x(t_0))} g(\x(t_0)) \nonumber\\
  &\quad + \sum_{k=0}^{\infty} \frac{(\beta (t\!-\!t_0))^{k+1}}{(k\!+\!1)!}\lt[\A\diag(\bm{1} - \x(t_0))\rt]^k \A b(\x(t_0)), \label{upper-bound}
\end{align}
and $\xx(t)$ is given by (\ref{basic-SI-linear-solution}). In addition, $\lVert\xxx(t)-\x(t)\rVert$ tends to zero while $\lVert\xx(t)-\x(t)\rVert$ tends to infinity, as time $t$ goes to infinity.
\end{theorem}
\vspace{-3mm}
\begin{proof}
See Appendix~\ref{thm-upper-proof}.
\end{proof}
\vspace{-2mm}
\vspace{-0mm}
\begin{remark}
Theorem~\ref{thm:upper} shows that our upper bound $\xxx(t) \!=\! f(\yyy(t))$ is provided as a function of $\yyy(t)$ whose closed-form expression is given by (\ref{upper-bound}) with the initial condition $\yyy(t_0) \!=\! g(\x(t_0))$. The diagonal matrix $\diag(\bm{1} - \x(t_0))$ has diagonal elements $1 - x_i(t_0)$. Similarly, $b(\x(t_0))$ has entries $x_i(t_0) + (1\!-\! x_i(t_0))\log(1\!-\! x_i(t_0))$. Therefore, with a given snapshot $\x(t_0)$ at $t_0$ and given $\beta, \A$, one can readily evaluate $\xxx(t)$ for $t \!\geq\! t_0$, where the power-series term can be evaluated by taking a truncation of the series.\footnote{We can show that $\lVert [\A\diag(\bm{1} \!-\! \x(0))]^k \A \x(0) \rVert \lessapprox [\lambda(\A\diag(\bm{1} \!-\! \x(0)))]^k \lambda(\A)\lVert\x(0)\rVert$, where $\lambda(\A\diag(\bm{1} \!-\! \x(0)))$ and $\lambda(\A)$ are the spectral radiuses of $\A\diag(\bm{1} \!-\! \x(0))$ and $\A$, respectively. By Stirling's approximation, $k! \!\sim\! \sqrt{2\pi k}(k/e)^k$, for given $\beta, t$, the $k$-th summand decays fast in $k$. The decaying speed is even faster for early $t$.} In addition, as shown in (\ref{ordering}) in Theorem~\ref{thm:upper}, our bound $\xxx(t)$ is tighter than the traditional linearization bound $\xx(t)$ for all time $t$. In particular, $\xxx(t)$ works for any choice of $\x(t_0)$ and remains valid as a probability for all $t$ (since $\hat{x}_i(t) \!=\! f(\hat{y}_i(t)) \!\in\! [0,1]$ for all $t$), in contrast to the linearization bound $\xx(t)$ that grows exponentially fast without bound.
\end{remark}

The overall procedure of our transformation method to obtain $\xxx(t)$ as an accurate approximation to $\x(t)$ can be summarized in the following steps: \textbf{1)} transform $\x(t)$ to $\y(t)$, \textbf{2)} bound $\y(t)$ by $\yyy(t)$, and \textbf{3)} inversely transform $\yyy(t)$ back to $\xxx(t)$. See the details in the proof of Theorem~\ref{thm:upper}. The whole procedure is non-trivial and more systematic as compared to the traditional, naive linearization method of obtaining $\xx(t)$. We also note that the closed-form expression of $\yyy(t)$ in (\ref{upper-bound}) can be simplified to the following.

\vspace{-0mm}
\begin{corollary}\label{cor}
Let $t_0 \!=\! 0$. If $\x(0) \prec \bm{1}$, (\ref{upper-bound}) reduces to
\begin{equation}
  \yyy(t) = g(\x(0)) + \lt[e^{\beta t \A\diag(\bm{1} - \x(0))} - \I\rt]\diag(\bm{1} \!-\! \x(0))^{-1}\x(0). \label{special1}
\end{equation}
If $x_i(0) \in \{0,1\}$ for all $i$, (\ref{upper-bound}) reduces to
\begin{equation}
  \yyy(t)  = g(\x(0)) + \sum_{k=0}^{\infty} \frac{(\beta t)^{k+1}}{(k\!+\!1)!}\lt[\A\diag(\bm{1} \!-\! \x(0))\rt]^k \A \x(0) \label{special2}
\end{equation}
\end{corollary}
\vspace{-2mm}
\begin{proof}
See Appendix~\ref{cor-proof}.
\end{proof}
\vspace{-0mm}

Corollary~\ref{cor} says that when $0 \!\leq\! x_i(0) \!<\! 1$ for all $i$, (\ref{upper-bound}) can be simplified to (\ref{special1}), which is essentially from the fact that $\diag(\bm{1} \!-\! \x(0))$ is invertible. Similarly, when $x_i(0) \!=\! 0$ or $x_i(0) \!=\! 1$ for all $i$, we have $\diag(\bm{1} \!-\! \x(0))g(\x(0)) \!=\! \bm{0}$ and $b(\x(0)) \!=\! \x(0)$, from which (\ref{upper-bound}) reduces to (\ref{special2}). See the proof of Corollary~\ref{cor} for more details. Note that for the other cases of $\x(0)$, (\ref{upper-bound}) can still be evaluated.

In addition, if $0 \!\leq\! x_i(0) \!<\! 1$ for all $i$, (\ref{special1}) can be further simplified as was done for $\xx(t)$ in (\ref{EVC-approx}) through the spectral decomposition of $\A$ and the power series of $e^{\A t}$. Let $\mu_1, \mu_2, \ldots, \mu_n$ be the eigenvalues of $\A\diag(\bm{1} - \x(0))$, with $\mu_1 \!=\! \lambda(\A\diag(\bm{1} - \x(0)))$. Let $\uuu_1$ and $\vvv_1$ be the left and right eigenvectors corresponding to $\mu_1$, normalized so that $\uuu_1^T\vvv_1 \!=\! 1$, respectively. We here assume that $\A\diag(\bm{1} \!-\! \x(0))$ is diagonalizable. Note that $\A$ is irreducible and aperiodic, and so is $\A\diag(\bm{1} \!-\! \x(0))$. Then, we have the following.
\begin{corollary}\label{cor2}
Let $t_0 \!=\! 0$. If $\x(0) \prec \bm{1}$, we have
\begin{align}
  \yyy(t) &= \hat{\xi}_1 e^{\beta \mu_1 t}\vvv_1\lt(1 + O\lt(e^{-\min_{k\geq 2}\!|\mu_1 - \mathrm{Re}(\mu_k)| t}\rt)\rt)  \nonumber \\
  &\quad - \diag(\bm{1} - \x(0))^{-1}\x(0) + g(\x(0)), \label{M-EVC-approx}
\end{align}
where $\hat{\xi}_1 \!\triangleq\! \uuu_1^T\diag(\bm{1} \!-\! \x(0))^{-1}\x(0) \!>\! 0$, and $\mu_1$ is real and positive.
\end{corollary}
\vspace{-2mm}
\begin{proof}
See Appendix~\ref{cor2-proof}.
\end{proof}
\vspace{-0mm}

As can be seen from (\ref{M-EVC-approx}), the growth rate of $\yyy(t)$ for \emph{large} time $t$ is mainly governed by the leading right eigenvector $\vvv_1$, which has a natural interpretation as the EVC of a \emph{weighted, directed} graph with its weight matrix $\W \!\triangleq\! \A\diag(\bm{1} \!-\! \x(0))$, where $w_{ij} \!=\! a_{ij} (1 \!-\! x_j(0))$ for all $i,j$. Note that $a_{ij} \!=\! 1$ should be read as a `directed' edge \emph{from} $j$ \emph{to} $i$, as can be seen from (\ref{basic-SIS}) and (\ref{basic-SI}). Despite the simplification and the natural interpretation, the dominant term involving the EVC of $\W$ in (\ref{M-EVC-approx}) alone may not be enough to accurately capture the SI epidemic dynamics, especially when the relevant timescale for the SI dynamics is the early time $t$.

\begin{figure}[t!]
    \subfigcapskip = -1mm
    \centering
    \vspace{-0mm}
    \hspace{-10mm}\subfigure[$\beta = 0.01$]{\includegraphics[width=1.65in,height=1.28in]{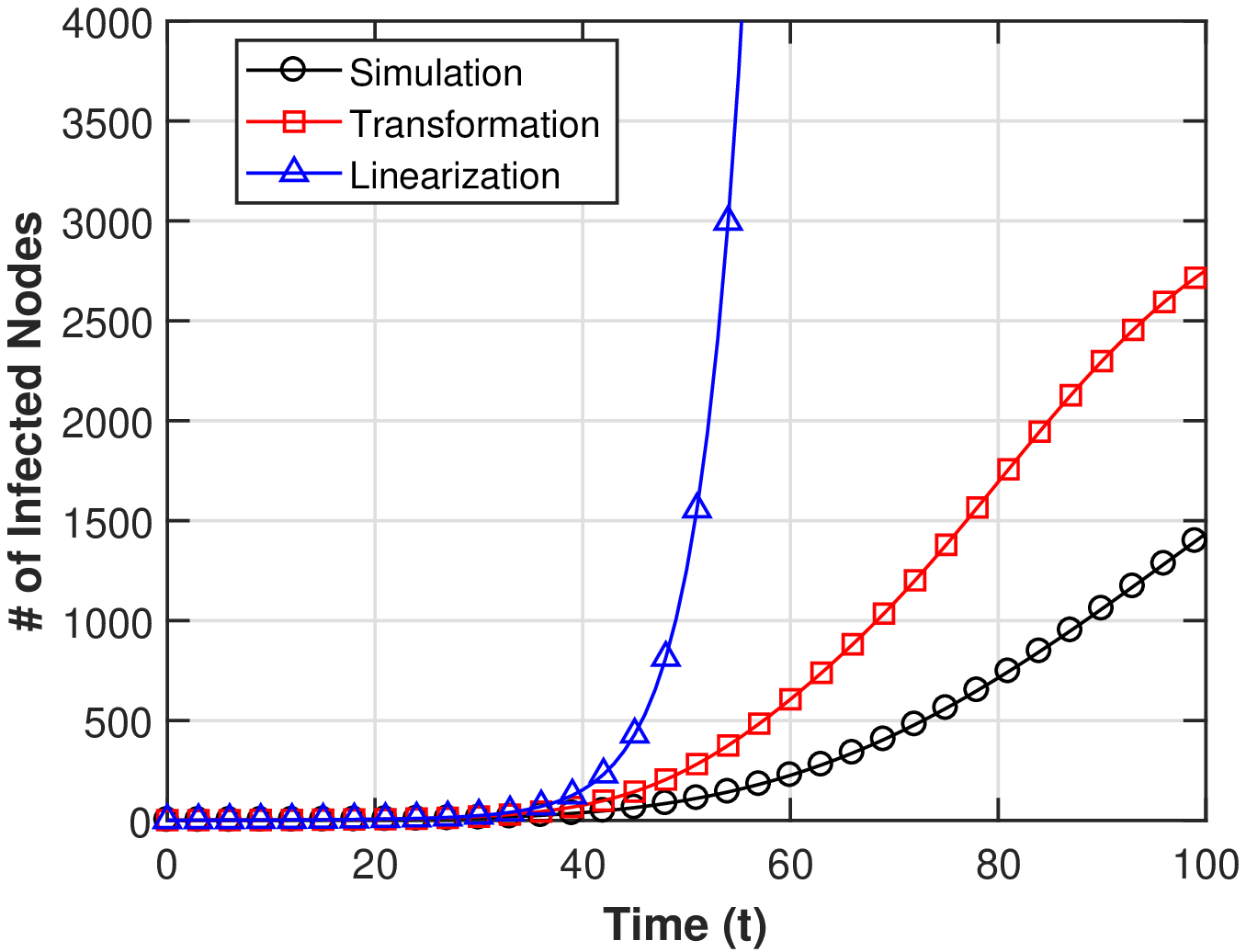}}
    \hspace{-0mm}\subfigure[$\beta = 0.05$]{\includegraphics[width=1.65in,height=1.28in]{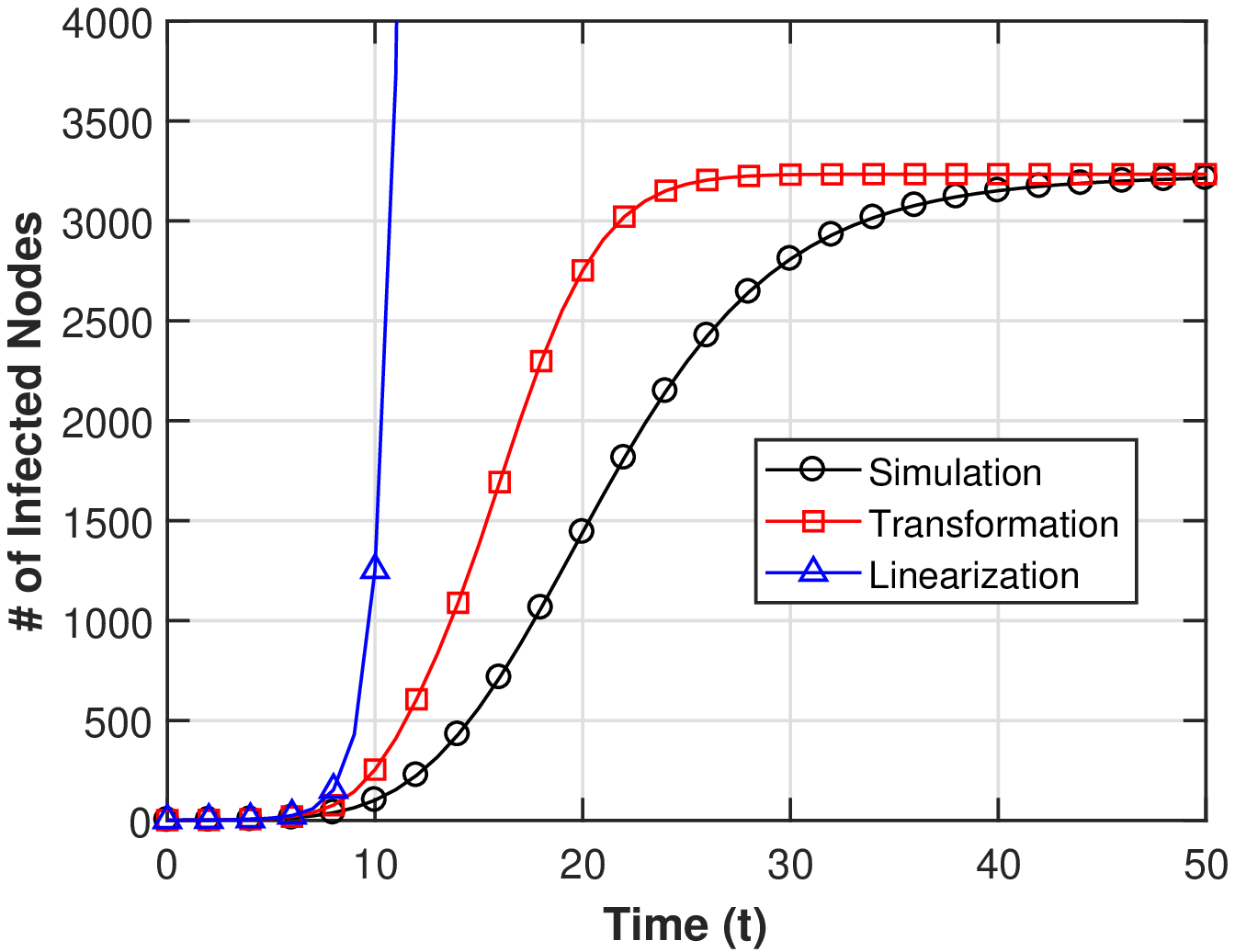}}
    \hspace{-12mm}
    \vspace{-4mm}
    \caption{\small The expected number of infected nodes $\sum_i x_i(t)$ (Simulation) obtained through the SI simulations and its upper bounds $\sum_i \tilde{x}_i(t)$ (Linearization) and $\sum_i \hat{x}_i(t)$ (Transformation) over time $t$.}\label{fig:comparison}
    \vspace{-4mm}
\end{figure}

We again emphasize that our bound $\xxx(t)$ (without the simplification) is valid for \emph{all} time $t$ and effectively overcomes the limitations of the linearization bound $\xx(t)$ in Section~\ref{se:model}, as shown in Theorem~\ref{thm:upper}. To numerically demonstrate the effectiveness of our bound $\xxx(t)$, in Figure~\ref{fig:comparison}, we provide the expected number of infected nodes $\sum_i x_i(t) \!=\! \Ex\lt\{\sum_i X_i(t)\rt\}$ obtained through the SI simulations, and its bounds $\sum_i \tilde{x}_i(t)$ (traditional bound) and $\sum_i \hat{x}_i(t)$ (our bound) over time $t$. Figure~\ref{fig:comparison} clearly demonstrates how quickly $\xx(t)$ becomes invalid, not to mention the huge gap between $\xx(t)$ and $\x(t)$, and confirms the superiority of $\xxx(t)$ over $\xx(t)$ to better approximate $\x(t)$ for all $t$. For Figure~\ref{fig:comparison}, we use the Gnutella P2P network dataset in the SNAP repository~\cite{snap}. Specifically, we use its largest strongly connected component (undirected version) with 3,234 nodes and 13,453 edges to ensure the connectivity. A fixed node is initially infected at time $0$. Our bound $\xxx(t)$ is computed based on the numerical evaluation of $\yyy(t)$ in (\ref{special2}) by taking a truncation of the series.

\vspace{-0mm}
\subsection{A New Look of the SI Model from Reliability Theory}

We turn our attention to a novel interpretation of the SI epidemic dynamics from reliability theory, and its connection to our transformation method and associated properties.

Without loss of generality, we set $t_0 \!=\! 0$. Let $T_i \!\triangleq\! \min\{t \!\geq\! 0: X_i(t) \!=\! 1\}$ be the time until node $i$ gets infected. While $T_i$ is a non-negative random variable, we allow that the event $\{T_i \!=\! 0\} \equiv \{X_i(0) \!=\! 1 \}$ is possible with non-zero probability $x_i(0) \!=\! \pr\{X_i(0) \!=\! 1 \} \!>\! 0$. We first observe the identity that
\begin{equation}
  x_i(t) = \Ex\{X_i(t)\} = \pr\{X_i(t) = 1\} = \pr\{T_i \leq t\},  \;\; i \in \N. \label{identity}
\end{equation}
Letting $f_{T_i}(t)$ be the probability density function of $T_i$, we have $dx_i(t)/dt \!=\! f_{T_i}(t)$ for all $i$. We also define by $h_i(t)$  the failure (or hazard) rate function~\cite{Ross96a,Lai2006}
\begin{equation}
  h_i(t) \triangleq \frac{f_{T_i}(t)}{\pr\{T_i>t\}} = \frac{\frac{d}{dt}\pr\{T_i\leq t\}}{\pr\{T_i>t\}},  \;\; i \in \N. \label{failure-rate-def}
\end{equation}
It is easy to see that with probability $h_i(t)dt$, node $i$ becomes infected in the interval $(t, t \!+\! dt)$ given that it has survived for time $t$. We then have the following.

\vspace{-0mm}
\begin{lemma} \label{lemma:survival}
The time-to-infection distribution is given by
\begin{equation}
  \pr\{T_i > t\} = \pr\{T_i > 0\}\exp\lt\{-\int_{0}^{t} h_i(s) ds\rt\}, \label{surv-func}
\end{equation}
with the failure rate function $h_i(t)$ given by
\begin{equation}
  h_i(t) = \beta \sum_{j\in\N} a_{ij}x_j(t) = \beta \sum_{j\in\N} a_{ij}\pr\{T_j \leq t\}, \;\; i \in \N. \label{failure-rate}
\end{equation}
\end{lemma}
\vspace{-3mm}
\begin{proof}
Using the identity in (\ref{identity}) and $\frac{dx_i(t)}{dt} \!=\! f_{T_i}(t)$, we observe that (\ref{basic-SI}) can be written as $f_{T_i}(t) \!=\! \pr\{T_i \!>\! t\} \beta \sum_{j\in\N} a_{ij} \pr\{T_j \!\leq\! t\}$,
which implies
\begin{equation*}
  h_i(t) = \frac{f_{T_i}(t)}{\pr\{T_i>t\}} = \beta \sum_{j\in\N} a_{ij}\pr\{T_j \leq t\}.
\end{equation*}
After integrating both sides of the first equality and exponentiating the both sides, we have $\pr\{T_i \!>\! t\}  \!=\! C \exp\lt\{-\int_{0}^{t} h_i(s) ds\rt\}$. Letting $t \!=\! 0$ yields $C \!=\! \pr\{T_i \!>\! 0\}$, and thus we have (\ref{surv-func}).
\end{proof}
\vspace{-0mm}

It is worth noting that the exponential term in the RHS of (\ref{surv-func}) is the usual form of the survival function, $\pr\{T_i > t\}$, in terms of the failure rate function in the reliability theory, which assumes that the probability of surviving past time 0 is 1~\cite{Ross96a,Lai2006}. Such an assumption is relaxed here to capture the possibility that some nodes are already infected at time $0$, and thus the term $\pr\{T_i \!>\! 0\}$ appears in the RHS of (\ref{surv-func}).

Lemma~\ref{lemma:survival} shows that the failure rate function $h_i(t)$ uniquely determines the distribution $\pr\{T_i \!>\! t\}$. This is generally true for any continuous random variables~\cite{Ross96a,Lai2006}. Furthermore, by the construction of our transformation in (\ref{transform-y})--(\ref{transform-SI}) and from (\ref{identity}), we see that for any $i \in \N$,
\begin{equation}\label{hazard}
  y_i(t) = -\log(1 - x_i(t)) = -\log\lt( \pr\{T_i > 0\} \rt) + \int_{0}^{t} h_i(s) ds,
\end{equation}
implying that $\frac{d y_i(t)}{dt} \!=\! h_i(t)$. Thus, considering the relationship between $x_i(t)$ and $\pr\{T_i \!>\! t\}$ in (\ref{identity}), one can observe that the structure of the one-to-one correspondence between $x_i(t)$ and $y_i(t)$ is identical to the one between the failure rate function $h_i(t)$ and the distribution $\pr\{T_i \!>\! t\}$.

In addition, let $T_{i,t}$ be the \emph{residual life} at time $t$ whose distribution is given by
\begin{equation}
  \!\!\pr\{T_{i,t} > t'\} = \pr\{T_i > t + t' | T_i > t\} = \exp\lt\{ - \!\int_{t}^{t+t'} \!\!h_i(s) ds \rt\},\!\! \label{residual}
\end{equation}
from Lemma~\ref{lemma:survival}. We then have the following.
\vspace{-0mm}
\begin{lemma}\label{lemma:residual}
$T_{i,t}$ is \emph{stochastically} decreasing (non-increasing) in $t \geq 0$ for each $i$.
\end{lemma}
\vspace{-3mm}
\begin{proof}
Fix $i$. From (\ref{upper-bound}), we see that $y_i(t)$ is non-decreasing in $t$. Since $f(y)$ is monotone increasing in $y \!\geq\! 0$, this implies that the RHS of (\ref{transform-SI}) is non-decreasing, and so is $\frac{dy_i(t)}{dt}$. The identity $\frac{d y_i(t)}{dt} \!=\! h_i(t)$ implies that $h_i(t)$ is also non-decreasing in $t$. Thus, the assertion follows by noting the identity of $\pr\{T_{i,t} \!>\! t'\}$ in (\ref{residual}).
\end{proof}
\vspace{-0mm}

This means that the older node $i$ survives without being infected, the stochastically smaller its residual life to infection.

\vspace{-0mm}
\section{Combating SI Epidemic Spreading with Limited Resources}\label{se:application}

As noted in Section~\ref{se:practical}, there is a non-trivial technical challenge for combating epidemic spreading due to the limited resources -- both time and bandwidth required. It naturally translates into how to prioritize nodes for patching and vaccination under resource-constrained environments. We can broadly classify vaccination strategies into two categories based on the target scenarios. The first one is for \emph{proactive} or \emph{preventive} vaccination. In this case, no one has been infected by $t_0$. No one would know which kind of malware/worm attack would be a threat later on, but one can put forth an effort to proactively make the networking system more robust, for example, by utilizing the diversification of operating systems and software versions across nodes~\cite{Garcia11,Newell15}. In most relevant scenarios, due to limited resources, we are often asked to prioritize which nodes to be treated or patched first. This boils down to the problem of choosing a limited set of nodes for early treatment to maximize the return on spending limited resources, or equivalently, to minimize the potential impact from any possible epidemic outbreak in the future.

The second category is for \emph{reactive} vaccination. In this case, the spread of an epidemic already took place over the large-scale network. Assuming that the knowledge of such an on-going epidemic is available and can be estimated, the problem is how to mitigate the epidemic spreading to the extent possible with any resources available by $t_0$, which may be lately available patches or just incomplete workaround, e.g., changing the operating system. It becomes particularly important under cost-constrained networking environments, where immediate post-treatment/-patching for vulnerable (not-yet-infected) nodes is not possible. For example, the network may consist of many battery-constrained and/or resource-limited devices that frequently need to go into a dormant state, as would be the case in future large-scale IoT networks. It again boils down to the problem of choosing a limited set of nodes for such reactive vaccination so as to minimize the damage from the epidemic spread, now with a possible knowledge of highly contagious area or the estimated source of infection.

Let $K$ be a given number of patches/vaccines to immunize $K$ nodes over $\GG$. We are interested in designing vaccination policies to find which set of nodes need to be vaccinated out of $\N$ for the $K$ patches/vaccines in both preventively and reactively. The difference is that in the former, the source of a (future) epidemic outbreak can potentially be any node, i.e., $x_i(t_0) \!=\! c/n$, while the latter has a specific source of the epidemic outbreak, i.e., $x_i(t_0) \!=\! 1$ for some $i$. We first note that from the monotonicity property $\x(t_1) \preceq \x(t_2)$ for $t_1 \!\leq\! t_2$, the likelihood or probability $x_i(t)$ of node $i$ being infected increases in $t$. Lemma~\ref{lemma:residual} also says that the remaining lifetime (time till infection) of susceptible (not-yet-infected) node $i$ at time $t$ is stochastically decreasing in $t$. In other words, this node is more likely to be infected as time goes on, if not `treated' (or vaccinated) now. Thus, in order to maximally suppress down the (early) growth rate of the SI dynamics out of $K$ vaccines, a rule of thumb would be to find the most likely $K$ nodes who would become \emph{first} infected and to immunize these $K$ nodes.

If $\x(t)$ is available in a closed form, we would evaluate $\x(t)$ for early time $t$, sort $x_i(t), i \!=\! 1,2,\ldots,n$, in a descending order, and then choose the top $K$ nodes for vaccination. While $\x(t)$ is generally not known in a closed form, thanks to Theorem~\ref{thm:upper}, we are here able to leverage the closed-form expression $\xxx(t) \!=\! f(\yyy(t))$ with $\yyy(t)$ in (\ref{upper-bound}). That is, our vaccination policy is to sort $\hat{x}_i(t)$, or equivalently $\hat{y}_i(t)$ (since $f(y)$ is monotone increasing in $y$), $i \!=\! 1,2,\ldots,n$, in a decreasing order and to choose the corresponding top $K$ nodes for vaccine distribution. Since the prediction $\xxx(t)$ for $t \!\geq\! t_0$ via (\ref{upper-bound}) evolves from our initial `knowledge' $\xxx(t_0) \!=\! \x(t_0)$, our vaccine distribution strategies largely depend on the type of the initial knowledge and are categorized into the following preventive and reactive vaccination policies.

\vspace{1mm}
\noindent \textbf{Preventive Vaccination:} We set $t_0 \!=\! 0$ for ease of explanation. In this scenario, $x_i(0) \!=\! c/n$, $i\!=\! 1,2,\ldots, n$, for some $c$. We define a constant $\alpha \triangleq 1 \!-\! c/n$ for notational simplicity. By noticing $\diag(\bm{1} \!-\! \x(0)) \!=\! \alpha\I$, from Corollary~\ref{cor}, we have
\begin{equation}
 \yyy(t) = (1/\alpha - 1) e^{\alpha\beta t \A}\bm{1} - \lt(1/\alpha - 1 + \log(\alpha)\rt) \bm{1}, \label{preventive}
\end{equation}
where $e^{\alpha\beta t \A}$ can be written as
\begin{equation}
   e^{\alpha\beta t \A} = \I + \alpha\beta t\A + \frac{(\alpha\beta t)^2}{2!}\A^2 + \cdots \label{preventive2}
\end{equation}
Since $[\A^k]_{ij}$ gives the number of walks of length $k$ that connect nodes $i$ and $j$, $[e^{\alpha\beta t \A}]_{ij}$ indicates the \emph{weighted} sum of walks connecting nodes $i$ and $j$, where the number of length-$k$ walks is discounted by the weight $(\alpha\beta t)^k/k!$, putting relatively larger weights on shorter walks than longer ones. In other words, $[e^{\alpha\beta t \A}]_{ij}$ reflects how `easy' node $j$ can propagate a virus over $\GG$ to infect node $i$, and vice versa due to the symmetry of $e^{\alpha\beta t \A}$. Thus, if node $i$ has a higher weighted sum of walks from every node to itself than that of node $j$ at time $t$, i.e.,
\begin{equation*}
  [e^{\alpha\beta t \A}\bm{1}]_i = \sum_l [e^{\alpha\beta t \A}]_{il} \geq \sum_l [e^{\alpha\beta t \A}]_{jl}  = [e^{\alpha\beta t \A}\bm{1}]_j,
\end{equation*}
then node $i$ has a higher chance of being infected at $t$ than node $j$. That is, node $i$ has many relatively smaller paths of being infected from others at $t$. In addition, since $e^{\alpha\beta t \A}\bm{1}$ is the only dominant factor and is irrespective of $\x(0) = (c/n)\bm{1}$, this preventive vaccination is source-independent/agonistic.

It is worth noting that the $(i,j)$ entry $[e^{\A}]_{ij}$ of the matrix exponential $e^{\A}$ has been proposed in the statistical physics as a measure of the ``communicability"~\cite{Estrada08} between nodes $i$ and $j$ over $\GG$, which is a generalization of the shortest path. In this view, we can similarly interpret $[e^{\alpha\beta t \A}]_{ij}$ in (\ref{preventive2}) as a measure of the ``infectivity" from $j$ to $i$ at time $t$ when $\x(0) = (c/n)\bm{1}$.

\vspace{1mm}
\noindent \textbf{Reactive Vaccination:} We next move on to the scenario that the knowledge of the initial sources for the currently prevalent epidemic spreading is available. That is, letting $\Ii \triangleq \{i\!\in\! \N : x_i(0) \!=\! 1\}$, we have the knowledge of $\Ii$ at time $0$.

We first start with the case that $x_i(0) \!=\! 0$ for $i \in \Ss \triangleq \N \setminus \Ii$, i.e., we know whether or not each node is infected at $t_0 \!=\! 0$. The corresponding expression of $\yyy(t)$ is given by (\ref{special2}) in Corollary~\ref{cor}. We see that $g(x_i(0)) = \infty$ for all $i \in \Ii$ and $g(x_i(0)) = 0$ for all $i \in \Ss$, which indicates the first term in the RHS of (\ref{special2}) is merely an indicator of the sources. Thus, the second term in the RHS of (\ref{special2}) mainly decides the value of $\hat{y}_i(t)$ for $i \!\in\! \Ss$. Here, $[(\A\diag(\bm{1} \!-\! \x(0)))^k]_{ij}$ counts the number of $k$-length walks from $j$ to $i$ over $\GG$, but with removal of the (directed) edges emanating from the sources, which can be seen from that $[\A\diag(\bm{1} \!-\! \x(0))]_{ij} \!=\! 0$ for all $j \!\in\! \Ii$. Furthermore, $[\A \x(0)]_l \!\geq\! 1$ for all direct (one-hop) neighbors $l$ of $\Ii$, while  $[\A \x(0)]_l \!=\! 0$ for all the other nodes. Therefore, $\hat{y}_i(t)$, $i \!\in\! \Ss$, is still governed by the weighted sum of walks from all direct (one-hop) neighbors $l$ of $\Ii$ to $i$, albeit over the `altered' graph, where the number of length-$k$ walks from $l$ to $i$ is penalized by slightly different weights $(\beta t)^{k+1}/(k\!+\!1)!$ but also scaled by $[\A \x(0)]_l$. In contrast to the preventive case, $\yyy(t)$ clearly depends on the network structure $\A$ as well as the source information $\x(0)$, which makes our reactive vaccination `source-aware'.

The other case of $x_i(0) \!>\! 0$ for some $i \!\in\! \Ss$ reflects the situation that in addition to $\Ii$, there are other nodes that are suspected to be already infected, although uncertain. The knowledge of $\Ii$ itself can also be incomplete, possibly with rough estimates on $\x(0)$. Nonetheless, in any case, we can leverage the closed-form expression of $\yyy(t)$ in (\ref{upper-bound}) with $t_0 \!=\! 0$. While we can obtain similar interpretations on $\yyy(t)$ as earlier, we omit them for brevity.

\vspace{-0mm}
\section{Simulation Results}

In this section, we present simulation results to demonstrate the efficacy of our proposed vaccination policies for both preventive and reactive scenarios. To this end, we consider two real-world network datasets. One is the Gnutella P2P graph~\cite{snap}, whose largest connected component (undirected version) with 3,234 nodes and 13,453 edges is used to ensure the connectivity. The other is the Oregon AS router graph of 2,504 nodes and 4,723 edges~\cite{Tong-TKDD16}. We set the infection rate $\beta \!=\! 0.05$. Each data point reported is obtained by averaging over $10^4$ independent simulations. Due to space constraint, we here present representative simulation results.

\begin{figure}[t!]
    \subfigcapskip = -1mm
    \centering
    \vspace{-0mm}
    \hspace{-0mm}\subfigure[Gnutella P2P graph; $K \!=\! 200$]{\includegraphics[width=1.6in,height=1.2in]{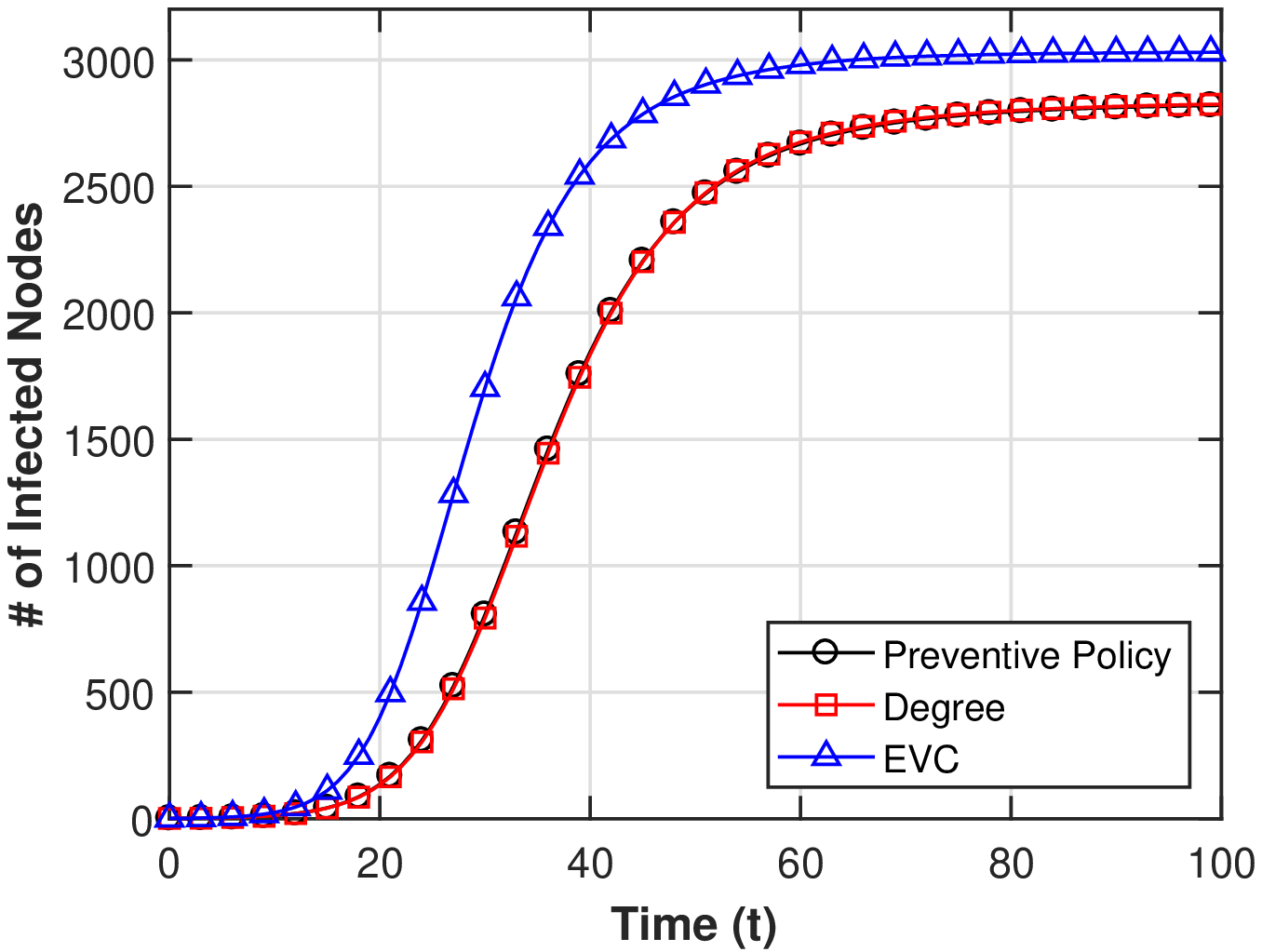}}
    \hspace{1mm}\subfigure[Gnutella P2P graph; $K \!=\! 500$]{\includegraphics[width=1.6in,height=1.2in]{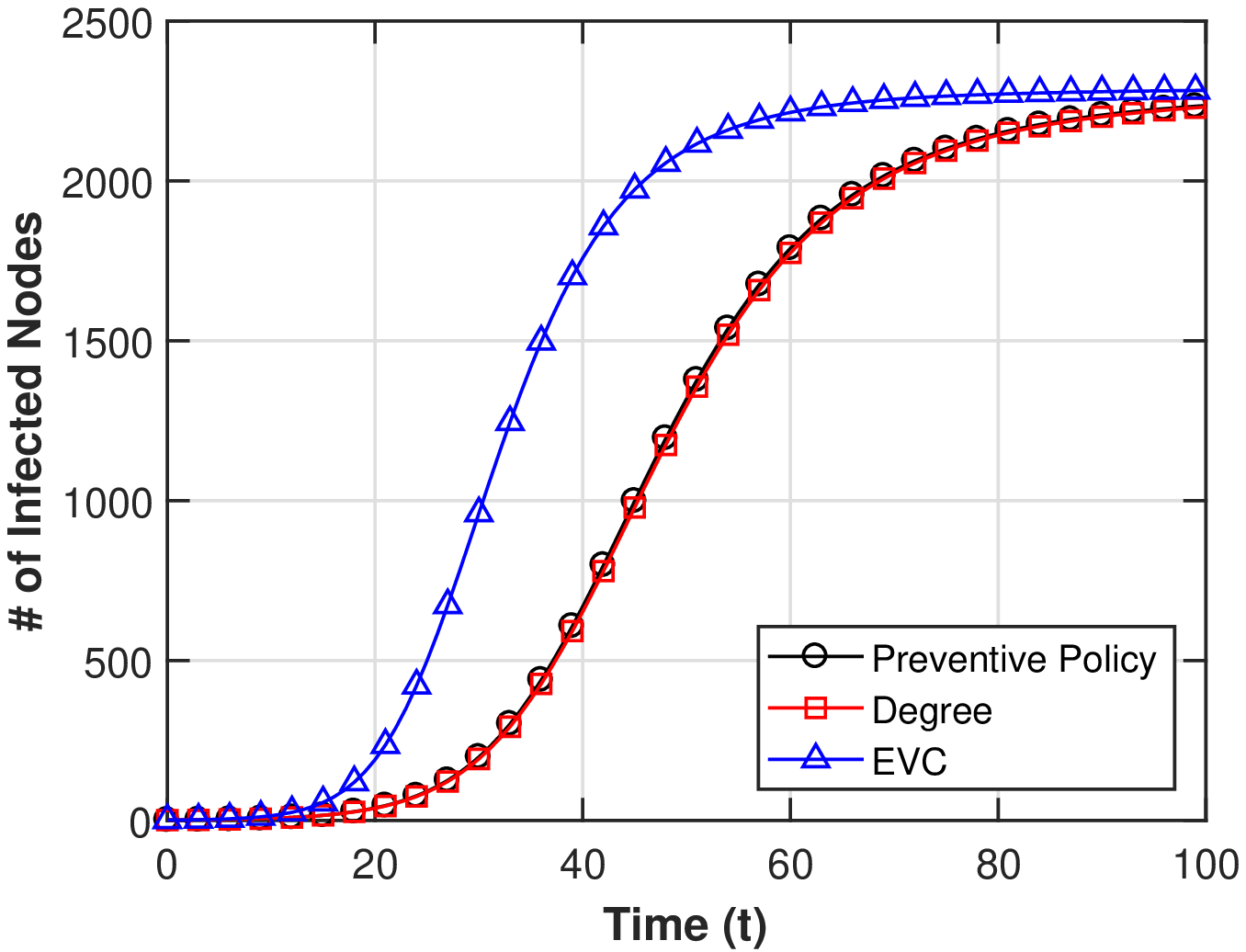}}
    \hspace{-0mm}\\
    \vspace{-1mm}
    \hspace{-0mm}\subfigure[Oregon AS graph; $K \!=\! 20$]{\includegraphics[width=1.6in,height=1.2in]{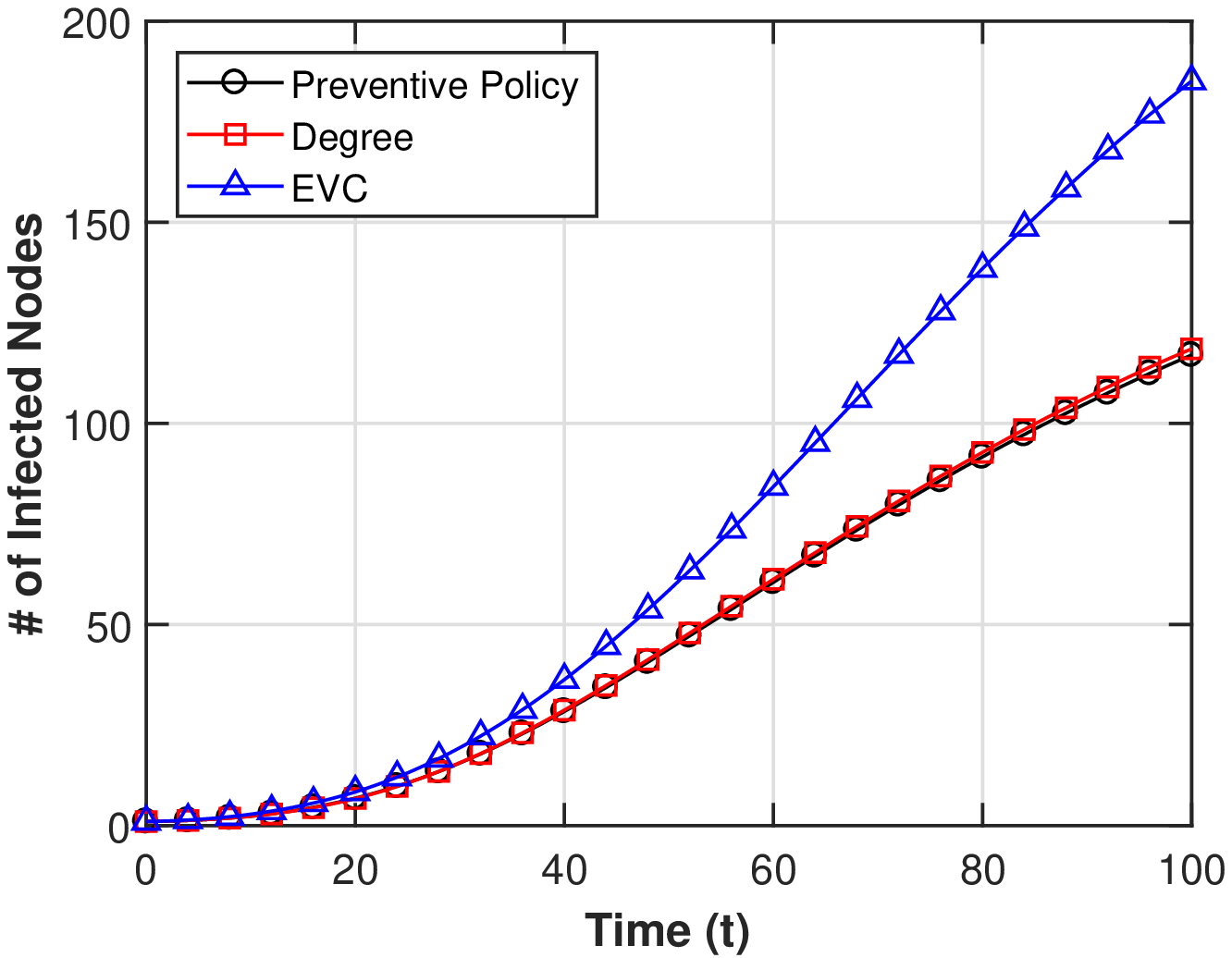}}
    \hspace{1.5mm}\subfigure[Oregon AS graph; $K \!=\! 30$]{\includegraphics[width=1.6in,height=1.2in]{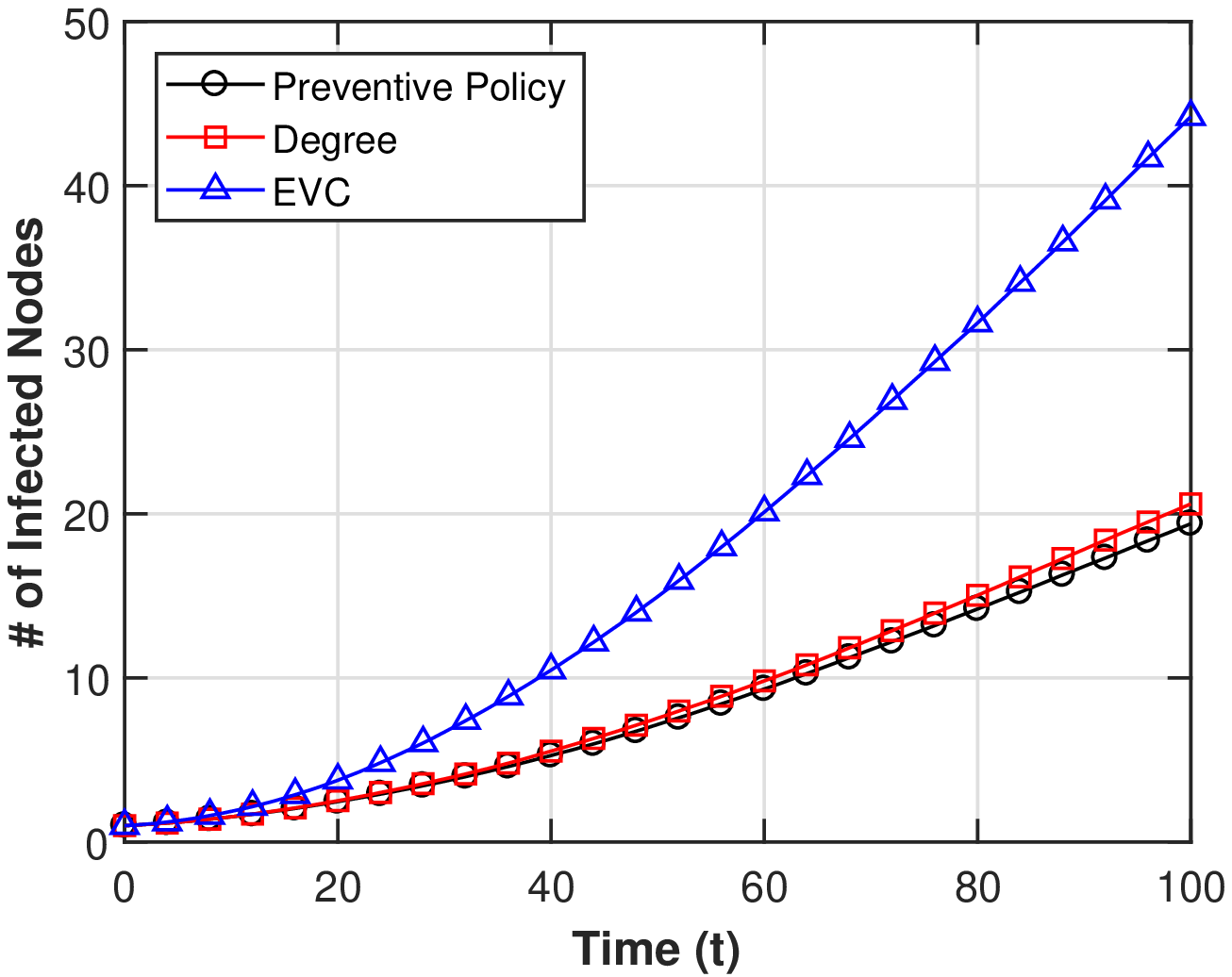}}
    \hspace{-0mm}
    \vspace{-4mm}
    \caption{\small \emph{Preventive scenario}. The expected number of infected nodes as time $t$ increases, using our preventive policy, EVC-based policy, and degree-based policy with different amounts of vaccines.}\label{fig:source-agnostic}
    \vspace{-3mm}
\end{figure}

We compare the performance of our vaccination policies against the EVC-based and degree-based policies. The former is the vaccination policy inspired by the conventional wisdom that the EVC can capture the early growth rate of the SI dynamics, which helps identify the critical nodes (or the most vulnerable nodes) for vaccine distribution~\cite{Canright06,Carreras2007,Newman10,Lu-PR16}. That is, this policy is to choose the top $K$ EVC nodes for immunization. The other policy is based on the node degree. From the graph connectivity, high-degree nodes have a high chance of being exposed to the spread of an epidemic and thus this policy is to choose the top $K$ degree nodes for vaccination. In all the policies, the selected nodes are fully immunized as if they are removed from the graph.

Figure~\ref{fig:source-agnostic} shows the expected number of infected nodes obtained over the Gnutella P2P and Oregon AS graphs for the preventive case, where an epidemic starts from a randomly chosen node, with different choices of $K$. In all cases, the EVC-based policy turns out to be the worst, which is in stark contrast to the conventional wisdom (built on top of the SI dynamics) that the node with high EVC value is considered to play a dominant role in spreading the infection over early time $t$. The unsatisfactory performance of the EVC-based policy is attributed to the third limitation of the linearization bound $\xx(t)$ mentioned in Section~\ref{se:model}. That is, the approximation of $\xx(t)$ via the EVC is only effective for \emph{large} time $t$ in which regime $\xx(t)$ already becomes invalid, i.e., $\tilde{x}_i(t) \!\gg\! 1$, and thus the EVC is not the right metric to capture the \emph{early} infection dynamics.

\begin{figure}[t!]
    \subfigcapskip = -1mm
    \centering
    \vspace{-0mm}
    \hspace{-0mm}\subfigure[Gnutella P2P graph; $K \!=\! 15$]{\includegraphics[width=1.6in,height=1.2in]{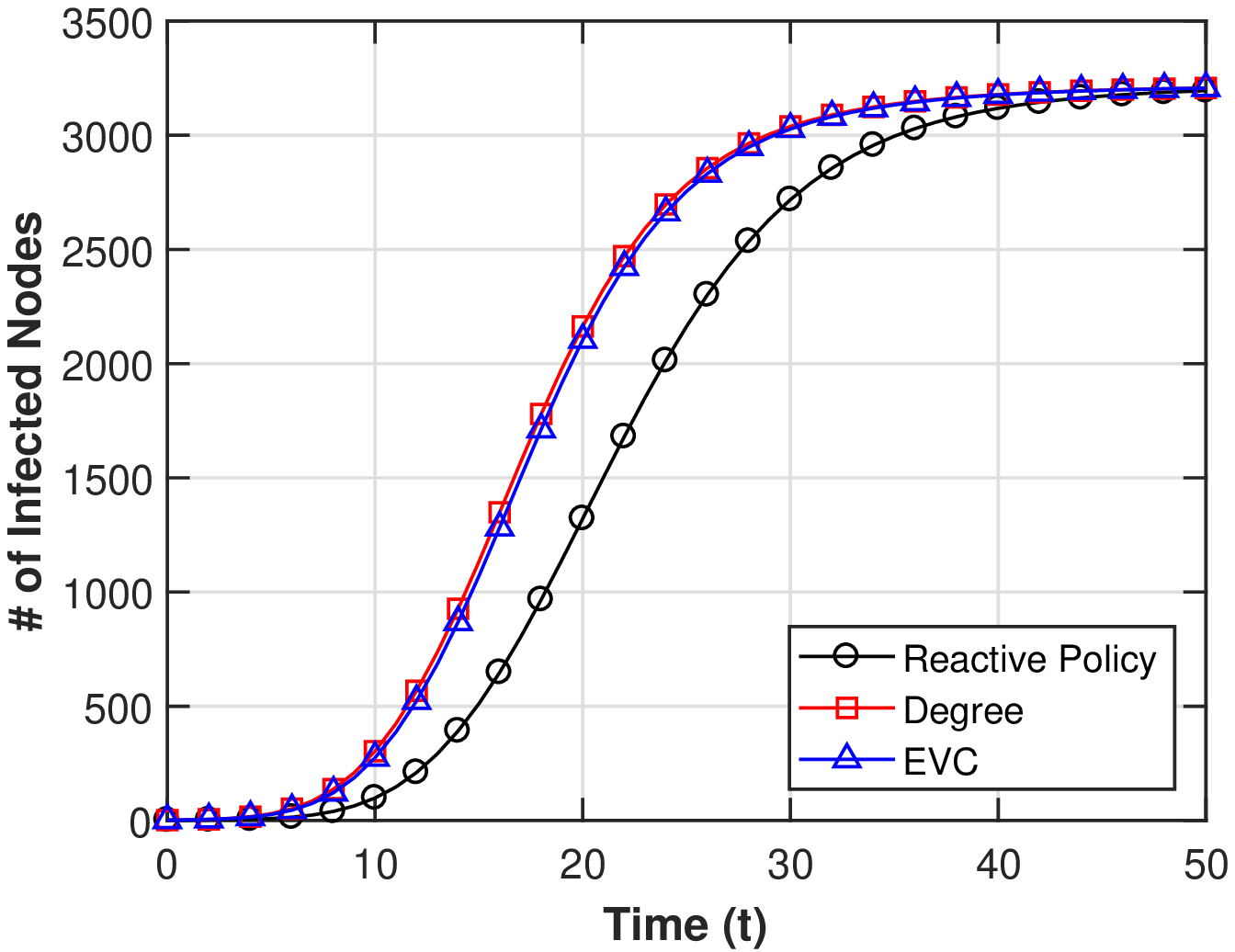}}
    \hspace{1mm}\subfigure[Gnutella P2P graph; $K \!=\! 18$]{\includegraphics[width=1.6in,height=1.2in]{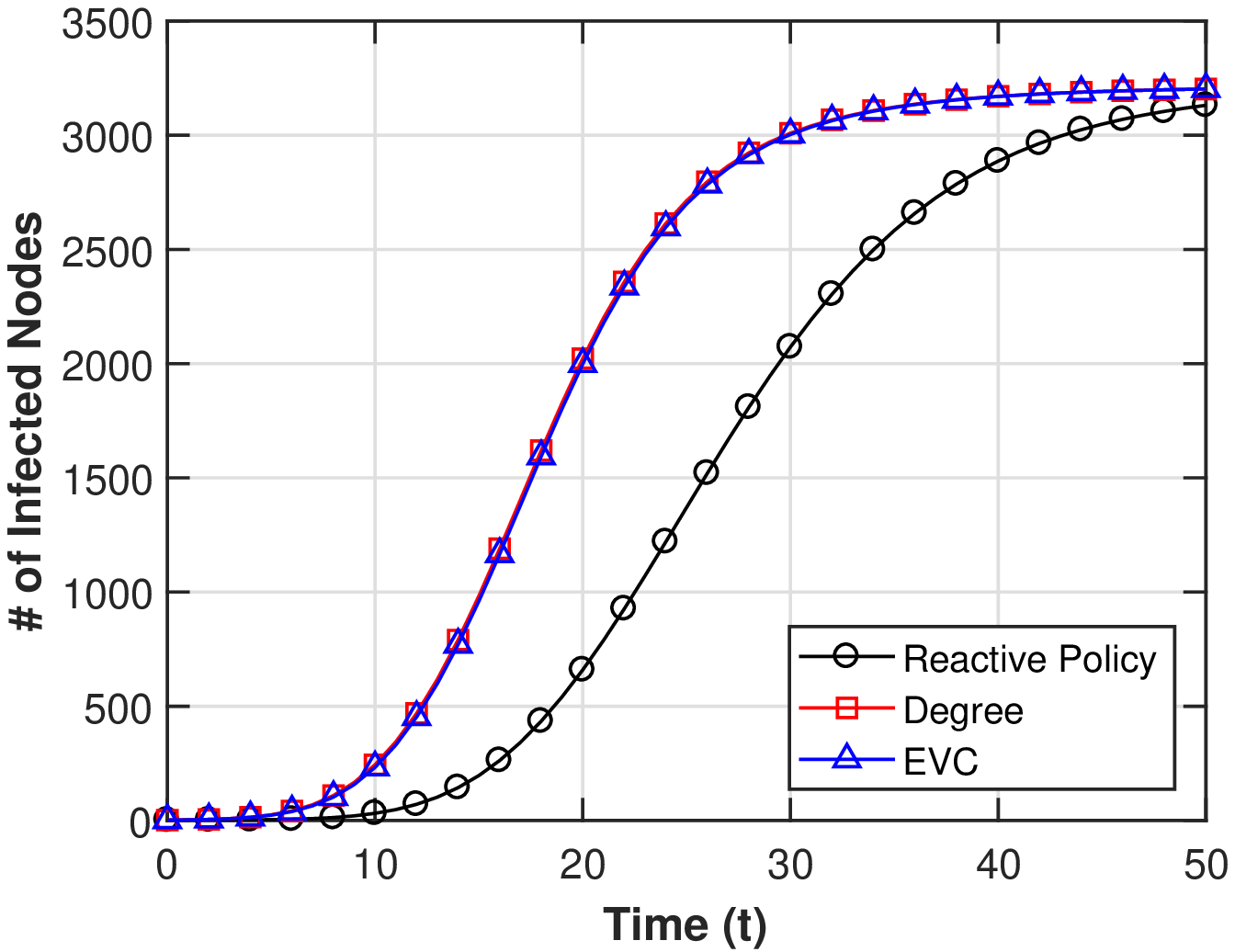}}
    \hspace{-0mm}\\
    \vspace{-1mm}
    \hspace{-0mm}\subfigure[Oregon AS graph; $K \!=\! 30$]{\includegraphics[width=1.6in,height=1.2in]{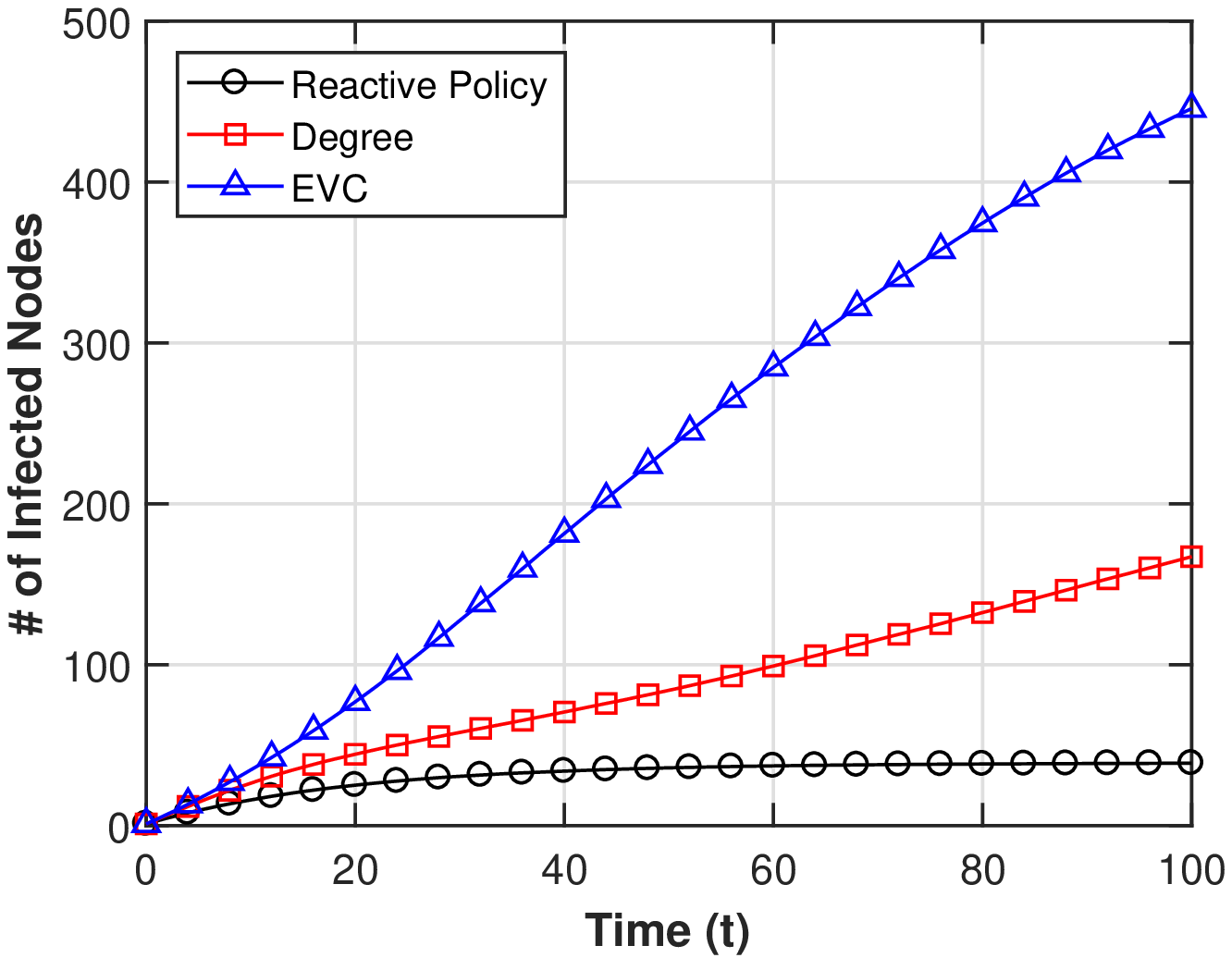}}
    \hspace{1.5mm}\subfigure[Oregon AS graph; $K \!=\! 50$]{\includegraphics[width=1.6in,height=1.2in]{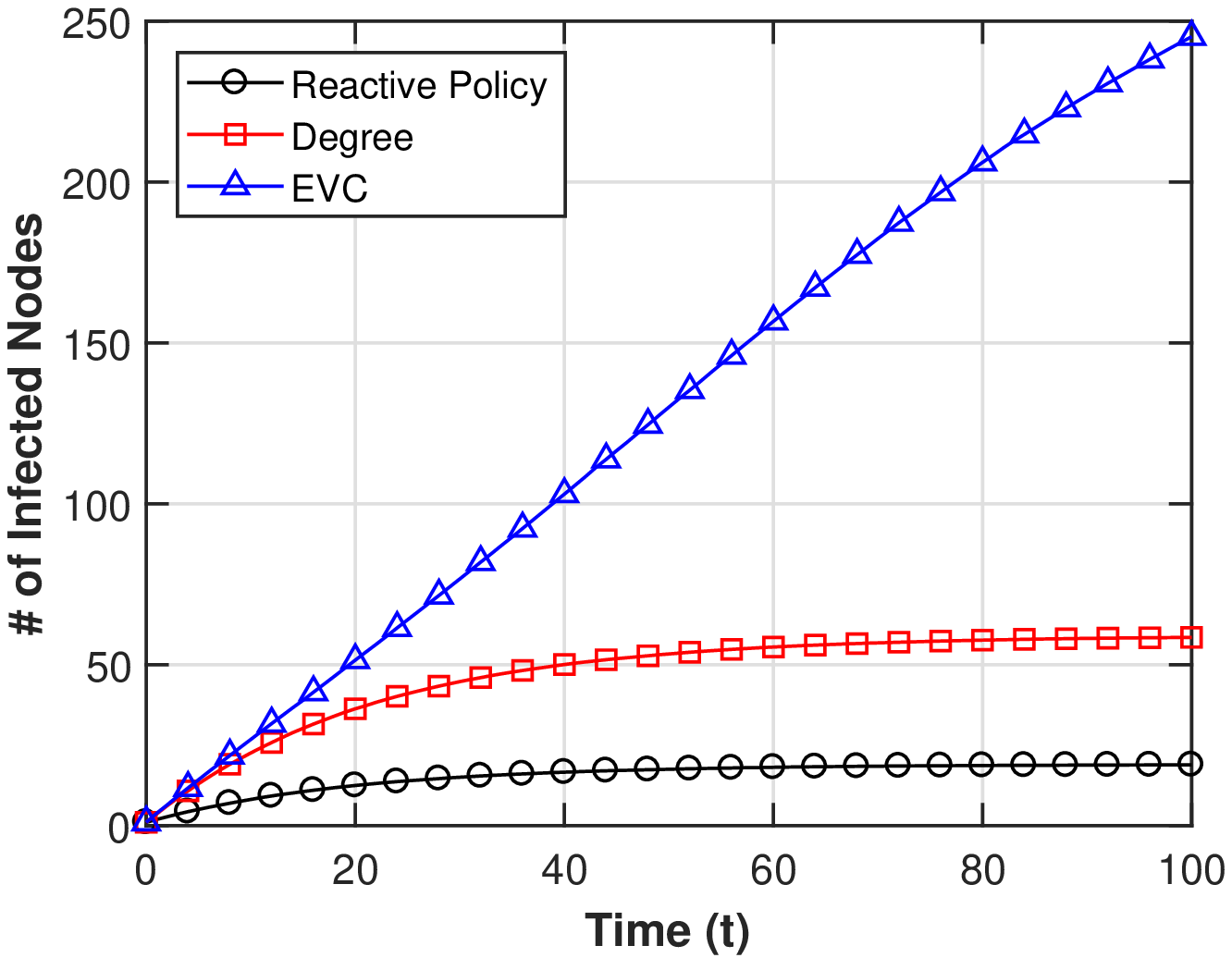}}
    \hspace{-0mm}
    \vspace{-4mm}
    \caption{\small \emph{Reactive scenario}. The expected number of infected nodes over time $t$ using our reactive policy, EVC-based policy, and degree-based policy with varying amounts of vaccines.}\label{fig:source-aware}
    \vspace{-3mm}
\end{figure}

On the other hand, our preventive policy and the degree-based policy lead to similar results, both of which are better than the EVC-based one. This is in line with our explanation under the notion of ``infectivity", implying that a node with many short-length walks of being connected from all other nodes is more likely to be selected in our preventive policy. Intuitively, high-degree nodes have a high chance of being connected to more one-hop or few-hop neighbors, and our preventive policy also gives higher weights to higher-degree nodes. To be precise, we note that the top K nodes under our preventive policy are purely determined by $e^{\alpha\beta t \A}\bm{1}$, as mentioned in Section~\ref{se:application}, and $e^{\alpha\beta t \A}\bm{1} \!\approx\! \bm{1} + \alpha\beta t\A \bm{1}$ for small $t$, as can be seen from (\ref{preventive2}). We also see that $\A \bm{1}$ is the vector of node degrees. Thus, they result in similar performance. It is worth noting that not all high-degree nodes have high EVC values, especially for those who are connected to low-degree nodes.

When it comes to the reactive scenario, our source-aware policy performs better than the other policies, as shown in Figure~\ref{fig:source-aware}, where a node with 22  neighbors is chosen to be the initial infective source for Gnutella P2P graph and a node with 68 neighbors is chosen for Oregon AS graph. Our policy clearly demonstrates the benefit of exploiting both the graph topology $\A$ and the currently available knowledge $\x(t_0)$, with limited resources available, i.e.,  $K \!=\! 15, 18$ for the Gnutella P2P graph and $K \!=\! 30, 50$ for the Oregon AS graph. We also observe that the performance improvement under the Oregon AS graph is more drastic. One possible reason is the structural difference between the graphs. The Gnutella P2P graph is a well-connected, random-like graph, while the Oregon AS graph has a highly skewed degree distribution. Our reactive policy, when combined with such a highly-skewed graph structure, can effectively \emph{fragment} the graph by vaccinating only few nodes with the knowledge of $\x(t_0)$, leaving a large portion of the graph no longer infected.

\section{Conclusion}

We have developed a theoretical framework to provide an accurate approximate solution to the standard SI epidemic model, which captures the temporal behavior of the SI dynamics over all time and is tighter than the existing linearized approximation. We also give a reliability-theory interpretation of the SI dynamics, which yields stochastic characterization of the residual life of each node until infection. The theoretical framework further enabled us to develop vaccination policies for both preventive and reactive mitigation of epidemic outbreaks. We believe that our work provides a first step toward the correct understanding of the transient dynamics of the SI epidemic spreading process, and also sheds light on the design of vaccination strategies to prioritize devices for vaccine distribution under resource-constrained environments.

\bibliographystyle{ACM-Reference-Format}
\bibliography{ref-all}

\appendix
\section{Proof of Theorem~\ref{thm:upper}}\label{thm-upper-proof}

We set $\x(t_0) = \xxx(t_0) = \xx(t_0)$. To show (\ref{ordering}), we first prove that
\begin{equation}
  \x(t) \preceq \xxx(t) = f(\yyy(t)), \;\; t \geq t_0,\label{part1}
\end{equation}
and $\yyy(t)$ is given by (\ref{upper-bound}). Observe that $\y(t) \!=\! g(\x(t))$ and $\yyy(t) \!=\! g(\xxx(t))$ are monotonic transformations of $\x(t)$ and $\xxx(t)$, respectively. It is thus enough to show that $\y(t) \preceq \yyy(t)$ when $\y(t_0) \!=\! \yyy(t_0)$. Considering the concavity of $f(y)$ and the tangent line of $f(y)$ at $y = y_0$, we have
\begin{equation*}
  f(y) \leq f(y_0) + f'(y_0) (y - y_0) = f'(y_0)y + f(y_0) - f'(y_0)y_0,
\end{equation*}
for $y \geq y_0 \geq 0$, where $f'(y) = e^{-y}$ is the derivative of $f(y)$. We also see that $y_i(t)$ is non-decreasing in $t \geq t_0$ with $y_i(t_0) \geq 0$ from the non-negativity of the RHS of (\ref{transform-SI}). Thus, we have
\begin{align}
  \frac{dy_i(t)}{dt} &= \beta \! \sum_{j\in\N} a_{ij}f(y_j(t)) \nonumber \\
  &\leq \beta \! \sum_{j\in\N} a_{ij} \lt[f'(y_j(t_0))y_j(t) \!+\! f(y_j(t_0)) \!-\! f'(y_j(t_0))y_j(t_0)\rt] \nonumber \\
  &= \beta \sum_{j \in \N} a_{ij} (1-x_j(t_0)) y_j(t) + \beta \sum_{j \in \N} a_{ij} b(x_j(t_0)), \label{ffff}
\end{align}
for $t \geq t_0$. The last equality follows from the identities   $f'(y_j(t_0)) = e^{-y_j(t_0)} = 1 - x_j(t_0)$, and
\begin{align*}
  f(y_j(t_0)) - f'(y_j(t_0))y_j(t_0) &= 1 - e^{-y_j(t_0)} (y_j(t_0) + 1) \\
  &= x_j(t_0) + (1\!-\!x_j(t_0)) \log(1\!-\!x_j(t_0)) \\
  &= b(x_j(t_0)),
\end{align*}
from $y_i(t_0) = -\log(1\!-\!x_i(t_0))$ and $b(x) = x + (1\!-\!x)\log(1\!-\!x)$. Then, we note that the RHS of (\ref{ffff}) with $y_j(t)$ replaced by $\hat{y}_j(t)$ in a matrix form is identical to the RHS of (\ref{upper2}). Using the method of variation of parameters~\cite{Braun93}, the solution $\yyy(t)$ of (\ref{upper2}) is given by
\begin{align}
  \yyy(t) &= e^{\beta(t-t_0) \A\diag(\bm{1} - \x(t_0))} g(\x(t_0)) \nonumber\\
  &\quad + \int_{t_0}^{t}\beta e^{\beta (t-s) \A\diag(\bm{1} - \x(t_0))} \A b(\x(t_0)) ds,  \;\; t \geq t_0. \label{ffff1}
\end{align}
By using the identity
\begin{equation*}
  e^{\beta s \A\diag(\bm{1} - \x(t_0))}  = \sum_{k=0}^{\infty} \frac{(\beta s)^k}{k!}\lt[\A\diag(\bm{1} \!-\! \x(t_0))\rt]^k,
\end{equation*}
the second term in the RHS of (\ref{ffff1}) can be written as
\begin{align}
  &\int_{t_0}^{t}\beta e^{\beta (t-s) \A\diag(\bm{1} - \x(t_0))} \A b(\x(t_0)) ds \nonumber \\
  &\quad = \sum_{k=0}^{\infty} \frac{\beta^{k+1}}{k!}\int_{t_0}^{t}(t\!-\!s)^k ds\lt[\A\diag(\bm{1} \!-\! \x(t_0))\rt]^k \A b(\x(t_0)) \nonumber \\
  &\quad =  \sum_{k=0}^{\infty} \frac{(\beta (t\!-\!t_0))^{k+1}}{(k\!+\!1)!}\lt[\A\diag(\bm{1} \!-\! \x(t_0))\rt]^k \A b(\x(t_0)). \label{ffff2}
\end{align}
Therefore, from (\ref{ffff}) and (\ref{ffff2}), we conclude that when $\y(t_0) = \yyy(t_0)$, $\y(t) \preceq \yyy(t)$, and thus $\x(t) \preceq \xxx(t)$ in (\ref{part1}), where $\yyy(t)$ is given by (\ref{upper-bound}). 

We next prove that
\begin{equation}
  \xxx(t) = f(\yyy(t)) \,\preceq\, \xx(t), \label{part2}
\end{equation}
for all $t \!\geq\! t_0$. Without loss of generality, we set $t_0 \!=\! 0$. To proceed, we need the following lemma.
\begin{lemma} \label{lemma-ordering}
If
\begin{equation}
  \frac{d\xxx(t)}{dt} = \frac{d f(\yyy(t))}{dt} \preceq \frac{d \yyy(t)}{dt} \preceq \frac{d\xx(t)}{dt}, \;\; t \geq 0,\label{suff}
\end{equation}
then (\ref{part2}) follows.
\end{lemma}
\vspace{-3mm}
\begin{proof}
Since $\xxx(t)$ and $\xx(t)$ start from the same initial points, i.e., $\xxx(0) \!=\! \xx(0) \!=\! \x(0)$, the ordering of their derivatives in (\ref{suff}) implies the ordering in (\ref{part2}) for all time $t \geq 0$.
\end{proof}

It remains to show (\ref{suff}), which is a stronger sufficient condition, to prove that (\ref{part2}) is satisfied. Together with (\ref{part1}), (\ref{part2}) implies  (\ref{ordering}). First, by noting $f(y) \!=\! 1 \!-\! e^{-y}$, observe that, for each $i$,
\begin{equation}
  \lt[\frac{d f(\yyy(t))}{dt} \rt]_i = e^{-\hat{y}_i(t)}\frac{d \hat{y}_i(t)}{dt} \leq  \frac{d \hat{y}_i(t)}{dt}, \label{suff1}
\end{equation}
for all $t \!\geq\! 0$, where the inequality follows from $e^{-y} \!\leq\! 1$ for all $y \!\in\! [0, \infty]$. Then, letting $\D \!\triangleq\! \diag(\bm{1} \!-\! \x(0))$, we write again $\yyy(t)$ in (\ref{upper-bound}) with $t_0 = 0$:
\begin{align*}
  \yyy(t) &= e^{\beta t \A\D} g(\x(0))  + \sum_{k=0}^{\infty} \frac{(\beta t)^{k+1}}{(k\!+\!1)!}(\A\D)^k \A b(\x(0)) \\
  &= \sum_{k=0}^{\infty} \frac{(\beta t)^k}{k!}(\A\D)^k g(\x(0)) + \sum_{k=0}^{\infty} \frac{(\beta t)^{k+1}}{(k\!+\!1)!}(\A\D)^k \A b(\x(0)),
\end{align*}
from the identity $e^{\beta t \A\D} = \sum_{k=0}^{\infty} \frac{(\beta t)^k}{k!}(\A\D)^k$. It then follows that
\begin{align}
  \frac{d\yyy(t)}{dt}  &= \beta \sum_{k=1}^{\infty} \frac{(\beta t)^{k-1}}{(k\!-\!1)!}(\A\D)^{k-1} \A\D  g(\x(0)) \nonumber \\
  &\quad + \beta \sum_{k=0}^{\infty} \frac{(\beta t)^k}{k!}(\A\D)^k \A b(\x(0)) \nonumber\\
  &= \beta \sum_{k=0}^{\infty} \frac{(\beta t)^k}{k!}(\A\D)^k \A \lt[\D  g(\x(0)) + b(\x(0))\rt] \nonumber\\
  &= \beta \sum_{k=0}^{\infty} \frac{(\beta t)^k}{k!}(\A\D)^k \A \x(0) = e^{\beta t \A\D} \beta \A \x(0), \label{ddd}
\end{align}
where the third equality follows from
\begin{align*}
  \lt[\D  g(\x(0)) + b(\x(0))\rt]_i &= -(1-x_i(0))\log(1-x_i(0)) \\
   &\quad + x_i(0) + (1-x_i(0))\log(1-x_i(0)), \\
   &= x_i(0), \;\; i = 1,2,\ldots, n,
\end{align*}
where $g(x) = -\log(1-x)$ and $b(x) = x + (1-x)\log(1-x)$. From (\ref{ddd}), we have
\begin{equation}
  \frac{d\yyy(t)}{dt} = e^{\beta t \A\D} \beta \A \x(0) \preceq e^{\beta \A t} \beta \A \x(0) = \frac{d\xx(t)}{dt}, \label{suff2}
\end{equation}
for all $t \!\geq\! 0$, where the inequality follows since all the elements of $\A$ and $\x(0)$ are non-negative with $\beta \!>\! 0$ and $1 \!-\! x_i(0) \leq 1$ on the diagonal of $\D$, and the last equality is from the fact that $\xx(t) = e^{\beta \A t}\x(0)$ as given in (\ref{basic-SI-linear-solution}). Therefore, (\ref{suff1}) and (\ref{suff2}) imply (\ref{suff}), and thus (\ref{part2}) follows by Lemma~\ref{lemma-ordering}. This completes the proof of (\ref{ordering}).

Lastly, recall the monotonicity property of $\x(t)$ over time $t$, i.e., $\x(t_1) \!\preceq\! \x(t_2)$ for any $t_1 \!\leq\! t_2$, and $\x(\infty) \!=\! \bm{1}$. From (\ref{upper-bound}), we see that $\hat{y}_i(t)$ is non-decreasing in time $t$ and grows without bound for all $i$. By $f(y) = 1 - e^{-y}$, it follows that $\lim_{t\to \infty} \hat{x}_i(t) = \lim_{t\to \infty} f(\hat{y}_i(t)) = 1$. In addition, from (\ref{basic-SI-linear-solution}), we have $\lim_{t\to \infty} \tilde{x}_i(t) = \infty$. Therefore, $\lVert\xxx(t) - \x(t) \rVert \to 0$ and $\lVert\xx(t) - \x(t)\rVert \to \infty$, as $t \to \infty$. This completes the rest of the proof.

\section{Proof of Corollary~\ref{cor}}\label{cor-proof}

For notational simplicity, let $\D \triangleq \diag(\bm{1} - \x(0))$. We first consider $0 \leq x_i(0) < 1$ for all $i$.  We write again $\yyy(t)$ in (\ref{upper-bound}) with $t_0 \!=\! 0$:
\begin{equation}
  \yyy(t)  = e^{\beta t \A\D} g(\x(0))  + \sum_{k=0}^{\infty} \frac{(\beta t)^{k+1}}{(k\!+\!1)!}(\A\D)^k \A b(\x(0)). \label{uuu}
\end{equation}
By noting that $\D$ is invertible since every diagonal entry of $\D$ is positive, i.e., $1 \!-\! x_i(0) \!>\! 0$, the second term in the RHS of (\ref{uuu}) can be written as
\begin{align}
  \sum_{k=0}^{\infty} \frac{(\beta t)^{k+1}}{(k\!+\!1)!}(\A\D)^k \A b(\x(0)) &= \sum_{k=0}^{\infty} \frac{(\beta t)^{k+1}}{(k\!+\!1)!}(\A\D)^k \A \D \D^{-1} b(\x(0))  \nonumber\\
  &= \sum_{k=0}^{\infty} \frac{(\beta t)^{k+1}}{(k\!+\!1)!}(\A\D)^{k+1} \D^{-1} b(\x(0)) \nonumber \\
  &= \lt[e^{\beta t \A\D} - \I \rt] \D^{-1} b(\x(0)), \label{uuu2}
\end{align}
with
\begin{equation*}
  \D^{-1} = \diag\lt((1\!-\!x_1(0))^{-1}, 1\!-\!x_2(0))^{-1}, \ldots, (1\!-\!x_n(0))^{-1}\rt).
\end{equation*}
Thus, from (\ref{uuu}) and (\ref{uuu2}), we have
\begin{align}
  \yyy(t) &= e^{\beta t \A\D}\lt[g(\x(0))+\D^{-1} b(\x(0))\rt] - \D^{-1} b(\x(0)) \nonumber\\
  &=  e^{\beta t \A\D}\D^{-1} \x(0) - \D^{-1} \x(0) + g(\x(0)) \nonumber\\
  &=  g(\x(0)) + \lt[e^{\beta t \A\D}  - \I \rt] \D^{-1} \x(0),  \label{uuu3}
\end{align}
from $g(x) = -\log(1-x)$ and $b(x) = x + (1-x)\log(1-x)$.

We next consider $x_i(0) \!=\! 0$ or $x_i(0) \!=\! 1$ for all $i$. Observe that the first term in the RHS of (\ref{upper-bound}) with $t_0 \!=\! 0$ can be simplified to
\begin{align}
   e^{\beta t \A\D} g(\x(0)) &= \sum_{k=0}^{\infty} \frac{(\beta t)^k}{k!}(\A\D)^k g(\x(0)) \nonumber \\
   &= \I g(\x(0)) + \sum_{k=1}^{\infty} \frac{(\beta t)^k}{k!}(\A\D)^{k-1}\A  \D g(\x(0)) \nonumber\\
   &= g(\x(0)), \label{uuu4}
\end{align}
where the last equality follows from $\D g(\x(0)) \!=\! \bm{0}$, since
\begin{equation*}
  \lt[\D g(\x(0))\rt]_i = -(1\!-\!x_i(0))\log(1\!-\!x_i(0)) = 0, \; \text{for all} \; i,
\end{equation*}
from $x_i(0) \!=\! 0$ or $x_i(0) \!=\! 1$ for all $i$. Thus, from (\ref{uuu4}),
$\yyy(t)$ in (\ref{upper-bound}) with $t_0 \!=\! 0$ can be written as
\begin{equation*}
  \yyy(t)  = g(\x(0)) + \sum_{k=0}^{\infty} \frac{(\beta t)^{k+1}}{(k\!+\!1)!}(\A\D)^k \A \x(0),
\end{equation*}
where we have also used the identity $b(\x(0)) = \x(0)$, i.e.,
\begin{equation*}
  b(x_i(0)) = x_i(0) + (1\!-\!x_i(0))\log(1\!-\!x_i(0)) = x_i(0), \; \text{for all} \; i,
\end{equation*}
from $x_i(0) \!=\! 0$ or $x_i(0) \!=\! 1$ for all $i$.

\section{Proof of Corollary~\ref{cor2}}\label{cor2-proof}

Consider $0 \!\leq\! x_i(0) \!<\! 1$ for all $i$. Letting $\D \triangleq \diag(\bm{1} - \x(0))$, we rewrite (\ref{special1}) as
\begin{equation}
  \yyy(t) = e^{\beta t \A\D}\D^{-1} \x(0) - \D^{-1} b(\x(0)). \label{eee}
\end{equation}
Recall that $\mu_1, \mu_2, \ldots, \mu_n$ are the eigenvalues of $\A\diag(\bm{1} - \x(0))$ and $\mu_1 \!=\! \lambda(\A\diag(\bm{1} - \x(0)))$. Let $\uuu_1, \uuu_2, \ldots, \uuu_n$ and $\vvv_1, \vvv_2, \ldots, \vvv_n$ be their associated left and right eigenvectors, respectively. They are here normalized so that $\uuu_i^T \vvv_j \!=\! \delta_{ij}$, where $\delta_{ij}$ is Kronecker delta~\cite{Bremaud99,Meyer00}.

First, observe that since $\A\D$ is diagonalizable, $e^{\beta t \A\D}$ can be decomposed as~\cite{Bremaud99,Meyer00}
\begin{equation}
  e^{\beta t \A\D} = \sum_{k=1}^n e^{\beta\mu_k t} \vvv_k \uuu_k^T. \label{eee1}
\end{equation}
Then, by noting that the non-negative matrix $\A\D$ is irreducible and aperiodic, the Perron-Frobenius theorem~\cite{Meyer00} implies that $\mu_1 \!=\! \lambda(\A\D)$ is positive with $\mu_1 \!>\! |\mu_k|$ ($k \!\neq\! 1$) and its corresponding left and right eigenvectors $\uuu_1$ and $\vvv_1$ have all positive components, i.e., $\uuu_1 \!\succ\! \bm{0}$ and $\vvv_1 \!\succ\! \bm{0}$, respectively. Thus, from (\ref{eee1}), the first term in the RHS of (\ref{eee}) can be written as
\begin{align}
  e^{\beta t \A\D}\D^{-1} \x(0) &= \sum_{k=1}^n e^{\beta\mu_k t} \vvv_k \uuu_k^T\D^{-1} \x(0) = \sum_{k=1}^n \hat{\xi}_k e^{\beta \mu_k t} \vvv_k \nonumber \\
  &= e^{\beta  \mu_1 t} \lt( \hat{\xi}_1 \vvv_1 + \sum_{k=2}^n\hat{\xi}_k e^{-\beta (\mu_1 - \mu_k) t} \vvv_k \rt), \label{eee2} \\
  &= \hat{\xi}_1 e^{\beta  \mu_1 t}\vvv_1 \lt(1 + O\lt(e^{- \min_{k \geq 2}\!|\mu_1 - \mathrm{Re}(\mu_k)| t}\rt)\rt), \label{eee3}
\end{align}
where $\hat{\xi}_k \!\triangleq\! \uuu_k^T\D^{-1} \x(0) \!\in\! \mathbb{R}$ for each $k$. By noting that $\mu_1 \!>\! 0$, $\vvv_1 \!\succ\! \bm{0}$, and $\hat{\xi}_1 \!=\! \uuu_1^T\D^{-1}\x(t_0) \!>\! 0$, (\ref{eee3}) follows from that all the summands in (\ref{eee2}) decay exponentially fast in time $t$, since $\mu_1 \!-\! \mu_k \!>\! 0$ for $k \!\geq\! 2$. It is worth noting that the eigenvalues $\mu_k$, $k \!\geq\! 2$, can possibly be complex-valued, as $\A\D$ is no longer symmetric. Nonetheless, since $\mu_1 \!-\! \mathrm{Re}(\mu_k) \!>\! 0$ for any $k \!\geq\! 2$ is preserved, the factor $e^{-\beta (\mu_1 - \mathrm{Re}(\mu_k)) t}$ still makes its corresponding summand exponentially decaying. Therefore, putting (\ref{eee3}) in (\ref{eee}), we are done.

\end{document}